\documentclass[%
 reprint,
 amsmath,amssymb,
 aps,longbibliography,
]{revtex4-1}

\usepackage{graphicx}
\usepackage{dcolumn}
\usepackage{bm}

\usepackage{xparse}

\usepackage{appendix}
\usepackage{graphicx}
\usepackage{dcolumn}
\usepackage{bm}

\newcommand{\ket}[1]{\ensuremath{\left| #1 \right>}}

\usepackage{bbold}

\begin{document}
\title{Exact Boundary Modes in an Interacting Quantum Wire}

\author{Colin Rylands}
 \email{crylands@umd.edu}
\affiliation{Joint Quantum Institute and
 Condensed Matter Theory Center, University of Maryland, College Park, MD 20742, USA}

\begin{abstract}
The boundary modes of one dimensional quantum systems can play host to a variety of remarkable phenomena. They can be used to describe the physics of impurities in higher dimensional systems, such as the ubiquitous Kondo effect or can support Majorana bound states which play a crucial role in the realm of quantum computation. In this work we examine the boundary modes in an interacting quantum wire with a proximity induced pairing term. We solve the system exactly by Bethe Ansatz and show that for certain boundary conditions the spectrum contains bound states localized about either edge.
The model is shown to exhibit a first order phase transition as a function of the interaction strength such that for attractive interactions the ground state has bound states at both ends of the wire while for repulsive interactions they are absent. In addition we see that the bound state energy lies within the gap for all values of the interaction strength but undergoes a sharp avoided level crossing for sufficiently strong interaction, thereby preventing its decay. This avoided crossing is shown to occur as a consequence of an exact self-duality which is present in the model. 

\end{abstract}
\date{\today}

\maketitle
\section{Introduction}

 Distinct equilibrium phases of matter are separated by regions of criticality characterized by diverging length scales and gapless modes. The critical region may lie in parameter space where the distinct phases are attained by tuning an external parameter such as temperature or some other system parameter such as an interaction strength\cite{DiFrancesco, Sachdev} but also occurs in real space at the interface of different systems. A prominent example of the latter is the existence of gapless modes which lie at the edges of topological materials\cite{Halperin, Hasan, Qi}. In one dimension, these edge modes are immobile and form bound states which are localized at the boundaries of the system. Such boundary bound states are of central importance to a number of fields including quantum computation\cite{Kitaev, Lutchyn, Oreg}, magnetic impurities in higher dimensional superconductors\cite{ Shiba, Rusinov, Yu, KondoSCRMP} and solitons in one dimensional organic conductors\cite{JackiwRebbi, SuShreiferHeeger, TakaLinMaki}.

In this paper we study a model of a one dimensional (1D), spinful, interacting quantum wire with open boundary conditions. The quantum wire has both density-density interactions and a proximity induced pairing term. We investigate the effects of different boundary conditions on the system and focus in particular on the existence of  boundary bound states, states which decay exponentially away from the endpoints of the wire. After reformulating the Hamiltonian via a bosonization and refermionization procedure the model is solved exactly using Bethe Ansatz for a subset of the parameters. We find the many body eigenstates, derive the Bethe Ansatz equations  and construct the ground state and low lying excitations. We show that for certain boundary conditions, those which break the time reversal invariance of the bulk, the model supports bound states at the boundaries. These boundary bound states, while being exact eigenstates of the system do not correspond to solutions of the Bethe Ansatz equations which marks them as distinct from boundary modes previously studied via Bethe Ansatz \cite{Wang, Saleur, Skorik, LeClair,  Ghoshal,GrisaruMezincescNepomechie,  ODBA, Grijalva}. 

When the interactions are absent these bound states lie within the energy gap at zero energy and provide a four-fold degenerate ground state. However when the interactions are present this degeneracy is lifted.  Owing to a redistribution of the interacting Fermi sea the bound state energy is shifted to non zero values leading to a first order  phase transition at zero temperature.  In the ground state these bound states are occupied if their energy is pushed below the Fermi level, which occurs for attractive interactions, and are unoccupied otherwise. The bound state energy remains within the gap for all values of the interaction strength however undergoes a sharp avoided level crossing with the continuum of unbound states for finite interaction strength. This prevents any possible coupling of the bound and unbound states and thereby protects them from decay\cite{Wigner}. 

\section{Hamiltonian}
We consider the following Hamiltonian describing an  interacting 1D quantum wire
\begin{eqnarray}\nonumber
H&=&\int^{L}_0 \mathrm{d}x\,\sum_{\sigma=\uparrow,\downarrow}v_F\left[\psi^\dag_{-,\sigma}i\partial_x\psi_{-,\sigma}-\psi^\dag_{+,\sigma}i\partial_x\psi_{+,\sigma}\right]\\\nonumber&&+\Delta\left[\psi^\dag_{+,\uparrow}\psi^\dag_{-,\downarrow}+\psi_{-,\downarrow}\psi_{+,\uparrow}-\psi^\dag_{+,\downarrow}\psi^\dag_{-,\uparrow}-\psi_{-,\uparrow}\psi_{+,\downarrow}\right]\\\nonumber&&+g_\parallel\left[\psi^\dag_{+,\uparrow}\psi_{+,\uparrow}\psi^\dag_{-,\uparrow}\psi_{-,\uparrow}+\psi^\dag_{-,\downarrow}\psi_{-,\downarrow}\psi^\dag_{+,\downarrow}\psi_{+,\downarrow}\right]\\
\label{H}&&+g_\perp\left[\psi^\dag_{+,\downarrow}\psi_{+,\downarrow}\psi^\dag_{-,\uparrow}\psi_{-,\uparrow}+\psi^\dag_{-,\downarrow}\psi_{-,\downarrow}\psi^\dag_{+,\uparrow}\psi_{+,\uparrow}\right].
\end{eqnarray}
Here we have two species, $\sigma=\uparrow,\downarrow$ of left ($-$) and right ($+$) moving fermions  $\psi^\dag_{\pm,\sigma}$, $\psi_{\pm,\sigma}$  which are restricted to the segment $x\in[0,L]$ and we have taken $\hbar=1$\cite{Giamarchi, GogolinNerseyanTsvelik, Tsvelik}. The first line is the kinetic energy of the fermions while the second is the pairing term which has a sign difference for pairs about different Fermi points.  Pairing of this form occurs in $p_x$-wave triplet superconductors\cite{Abrisokov,Sengupta} which may be induced via proximity \cite{ Lutchyn, Oreg,Sau,Fidkowski} and leads to a bare energy gap of $2\Delta$
. The final lines describe  density-density interactions between fermions with parallel  or opposite spins, the $g_\parallel$ and $g_\perp$ terms respectively. Along with this we specify the boundary condition at $x=0,L$ which changes the chirality of the particles but can be chosen to either conserve or flip the spin of the particle or some more complicated combination thereof. The first of these being the most natural choice in a quantum wire\cite{Fabrizio}. 

We shall see below that the spin conserving choice leads to the spectrum containing boundary bound states however the time reversal invariant, spin flipping boundary condition does not. These boundary bound states are invariant under a $\mathbb{Z}_2$, combined particle-hole and chirality transformation  $\psi^\dag_{\pm,\sigma}\leftrightarrow- \psi_{\mp,\sigma}$. This anti commutes with the non-interacting part of the Hamiltonian and so pins the bound states to zero energy when $g_\parallel=g_\perp=0$. In the interacting case however, we show below that this transformation  is generalized to  $H(\delta)\leftrightarrow- H(-\delta)$  where $\delta$ is the two particle phase shift and so the bound state may no longer lie at zero energy. 

\section{Bosonization}
Our aim is to provide an exact solution of the Hamiltonian, for a particular choice of $g_\perp,g_\parallel$, which will be achieved by employing the Bethe Ansatz method. The exact solution will then allow us to study the boundary bound states of the model. In present form however, \eqref{H} is not amenable to this due to the apparent lack of particle number conservation caused by the pairing term. To bring it to a more suitable form we first bosonize the system, perform a duality transformation and then refermionize. The outcome of this series of steps is that we will have a Hamiltonian in which the pairing term is replaced with a mass type term instead. Effectively this will be equivalent to a particle hole transformation for one of the chiral branches.

We introduce the bosonic fields $\psi^\dag_{\pm, \sigma}=\sqrt{D}e^{i[\mp\phi_{\sigma}-\theta_\sigma]}$ where $D=N/L$ is the average density\cite{Haldane}. Forming symmetric and antisymmetric combinations $\phi_\pm=[\phi_\uparrow\pm\phi_\downarrow]/\sqrt{2}$ and $\theta_\pm=[\theta_\uparrow\pm\theta_\downarrow]/\sqrt{2}$ which govern the charge ($+$) and spin ($-$) degrees of freedom our Hamiltonian becomes
\begin{eqnarray}\nonumber
H&=&\sum_{a=\pm}\frac{v_{a}}{2\pi}\int^{L}_0\mathrm{d}x \frac{1}{K_a}\left[\partial_x\phi_a(x)\right]^2+K_a\left[\partial_x\theta_a(x)\right]^2\\\label{Hbos}
&&-4\Delta D\sin[\sqrt{2}\phi_-]\sin[\sqrt{2}\theta_+]
\end{eqnarray}
where $v_{\pm}$ is the speed of sound and $K_\pm$ is the Luttinger parameter of the charge and spin components \cite{Giamarchi, GogolinNerseyanTsvelik, Tsvelik}. The relation between the fermionic parameters $g_\perp, g_\parallel$ and the bosonic parameters $K_\pm$  must be determined by comparing physical observables computed in both models and is non universal except at weak coupling wherein $K_\pm\approx 1-\left(g_\parallel\pm g_\perp\right)/2\pi v_F$. 

Next we make a duality transformation on the symmetric fields $\phi_+\leftrightarrow \theta_+$ 
whereupon we get the following 
\begin{eqnarray}\nonumber
H&=&\frac{v_{-}}{2\pi}\int^{L}_0\mathrm{d}x \frac{1}{K_-}\left[\partial_x\phi_-(x)\right]^2+K_-\left[\partial_x\theta_-(x)\right]^2\\\nonumber
&&+\frac{v_{+}}{2\pi}\int^{L}_0\mathrm{d}x K_+\left[\partial_x\phi_+(x)\right]^2+\frac{1}{K_+}\left[\partial_x\theta_+(x)\right]^2\\\label{Hbosdual}
&&-4\Delta D\int^{L}_0\mathrm{d}x\,\sin{[\sqrt{2}\phi_-]}\sin{[\sqrt{2}\phi_+]},
\end{eqnarray}
which is a variant of the double sine-Gordon model (DSG)\cite{Delfino, Fabrizio2}. The scaling dimension of the pairing term is $\varkappa=K_-/2+1/2K_+$ and it is known that for $\varkappa=1$ the DSG model is integrable \cite{Fateev} and enjoys a dual fermionic description\cite{BL,Lesage, SaleurBL}. We restrict ourselves to this case and define the new fermions $\mathcal{R}_{1,2}^\dag=\sqrt{D}e^{i[-\phi_{\uparrow,\downarrow}-\theta_{\uparrow,\downarrow}]}$ and  $\mathcal{L}_{1,2}^\dag=\sqrt{D}e^{i[-\phi_{\downarrow,\uparrow}-\theta_{\downarrow,\uparrow}]}$ in terms of which the Hamiltonian can be written as,
\begin{eqnarray}\nonumber
\mathcal{H}&=&\int^{L}_0\mathrm{d}x\,\sum_{\xi=1,2}v_F\left[\mathcal{L}_\xi^\dag i\partial_x\mathcal{L}_\xi-\mathcal{R}_\xi^\dag i\partial_x\mathcal{R}_\xi\right]\\\nonumber
&&+\Delta\left[\mathcal{R}^\dag_1\mathcal{L}_1+\mathcal{L}^\dag_1\mathcal{R}_1-\mathcal{R}^\dag_2\mathcal{L}_2-\mathcal{L}^\dag_2\mathcal{R}_2\right]\\\label{Hnew}
&&-2g\left[\mathcal{R}^\dag_1\mathcal{R}_1\mathcal{L}^\dag_2\mathcal{L}_2+\mathcal{R}^\dag_2\mathcal{R}_2\mathcal{L}^\dag_1\mathcal{L}_1\right].
\end{eqnarray}
Our new Hamiltonian, $ \mathcal{H}$, describes two new species of right moving, $\mathcal{R}_\xi^\dag, ~\mathcal{R}_\xi$ and left moving $\mathcal{L}_\xi^\dag, ~\mathcal{L}_\xi$ fermions which interact via density-density interaction and have mass term with $\Delta$ and $-\Delta$. Along our chosen manifold of $K_-+1/K_+=2$ there is only a single free parameter encoding the interactions in the wire which in $\mathcal{H}$ is given by $g$. The relationship between $g$ and $K_\pm$ can be  determined at this stage only at weak coupling $ 1/K_+-K_-\approx 2g/\pi v_F$. The full relationship will be discussed further below however we will refer to $g>0$ as the repulsive regime, $g<0$ as the attractive and $g=0$ being the non-interacting model, $K_+=K_-=1$.

As mentioned before this should be accompanied by boundary conditions at $x=0,L$. In terms of the new fermions the spin conserving boundary condition mixes the two species
\begin{eqnarray}\label{BC1}
\mathcal{R}_{1,2}^{\dag}(0)=-\mathcal{L}^\dag_{2,1}(0),\\\label{BC2}
\mathcal{R}_{1,2}^{\dag}(L)=-\mathcal{L}^\dag_{2,1}(L).
\end{eqnarray}
This will allow for boundary bound states to appear in the spectrum. The alternative choice which does conserves the species is given by 
\begin{eqnarray}\label{ABC1}
\mathcal{R}_{1,2}^{\dag}(0)=-\mathcal{L}^\dag_{1,2}(0),\\\label{ABC2}
\mathcal{R}_{1,2}^{\dag}(L)=-\mathcal{L}^\dag_{1,2}(L)
\end{eqnarray}
and  does not allow for boundary bound states to form. These boundary conditions are introduced as a point of comparison to the more natural and nontrivial case of \eqref{BC1}\eqref{BC2}, however it is worthwhile to note that it may be possible to engineer such boundary conditions through application of appropriate boundary fields or perhaps coupling to magnetic impurities\cite{Tim}.
 In both cases the minus sign accounts for the $\pi$ phase shift a particle acquires after scattering from a hard wall\cite{Fabrizio}.

\section{Boundary bound states}
We are now in a position to examine the system using Bethe Ansatz, first for the spin conserving boundary condition \eqref{BC1}. We begin by considering the single particle eigenstates and 
 introduce the notation $\Psi^\dag_\xi(x,\theta)=e^{i\Delta\sinh{(\theta)x/v_F}}\left[e^{\theta/2}\mathcal{R}_\xi^\dag(x)+\eta e^{-\theta/2}\mathcal{L}_\xi^\dag(x)\right]$ where $\eta=(-1)^{\xi-1}$, for $\xi=1,2$. In terms of this we can express the  single particle eigenstates as
\begin{eqnarray}\label{single}
\ket{\epsilon(\theta)}=\int_0^L\mathrm{d}x\sum_{\xi=1}^2\left[A_{+}^\xi\Psi^\dag_{\xi}(x,\theta)+A_{-}^\xi\Psi^\dag_{\xi}(x,-\theta)\right]\ket{0}\!.~
\end{eqnarray}
 Acting on this with the Hamiltonian we find that this has energy $\epsilon(\theta)=\Delta\cosh{(\theta)}$ and satisfies the boundary condition at $x=0$ provided $\vec{A}_+=K_L(\theta)\vec{A}_-$ where
\begin{eqnarray}\label{K}
K_L(\theta)=\frac{-1}{\cosh{(\theta)}}\begin{pmatrix}
1& -\sinh{(\theta)}\\
\sinh{(\theta)}& 1
\end{pmatrix}
\end{eqnarray}
and $\vec{A}_{\pm}=(A^1_{\pm},A^2_\pm)^T$. This also satisfies the boundary condition at $x=L$ if $e^{-2i\Delta \sinh{(\theta)}L/v_F}\vec{A}_-=K_R(\theta)\vec{A}_+$ where $K_R(\theta)=K^{-1}_L(\theta)=K_L(-\theta)$. Combining these two we get $e^{-2i\Delta \sinh{(\theta)}L/v_F}\vec{A}_-=K_R(\theta)K_L(\theta)\vec{A}_-
$
 which is the  boundary condition  of free fermions. The single particle rapidities therefore are quantized according to
 \begin{eqnarray}\label{BAEfree}
e^{-2i\Delta \sinh{(\theta)}L/v_F}=1.
\end{eqnarray} 
We may interpret this equation in the following fashion: The left hand side is the total phase shift accrued from the plane wave in \eqref{single} by a particle which travels a distance $2L$ from one side of the system to the other and then back to its original position. The right hand side provides the additional phase shift the particle acquires from scattering off of both boundaries. In this instance the contribution from both boundaries cancel each other. In the interacting case considered below, the right hand side will be significantly modified to account for the scattering between particles.

Solutions of this equation constitute single particle scattering states of the model and have a twofold degeneracy corresponding to the eigenvectors $\vec{A}_-=(1,0)^T~\text{or}~(0,1)^T$. Note that $\theta$ and $-\theta$ correspond to the same state while at $\theta=0$ the wavefunction vanishes. We therefore restrict to the real part of the rapidity  being positive, Re$(\theta)>0$ while the imaginary part may be zero or $\pi$, the latter choice giving negative energy particles. 

We may also construct zero energy boundary bound states of the model. Taking $\theta=i\pi/2$ and $A^\xi_-=0$ in the above expression for an eigenstate we have a state which decays as $e^{-\Delta x/v_F}$ and satisfies the boundary conditions at  both $x=0,L$ provided, $A_+^2=-iA_{+}^1$.  Explicitly, this is
\begin{eqnarray}\nonumber
\int_0^L\!\!\!\mathrm{d}x\,e^{-\Delta x/v_F}\Big\{\!\!\left[\mathcal{R}_1^\dag(x)-\mathcal{L}_2^\dag(x)\right]\!+\!i\left[\mathcal{R}_2^\dag(x)-\mathcal{L}_1^\dag(x)\right]\!\!\Big\}\!\ket{0}\!.
\end{eqnarray}
From this we can see that the bound state is invariant under the previously mentioned $\mathbb{Z}_2$ transformation, $\mathcal{R}^\dag_{1,2}(x)\leftrightarrow -\mathcal{L}^\dag_{2,1}(x)$ which preserves the boundary conditions \eqref{BC1}, \eqref{BC2}. The same transformation performed on the unbound state, \eqref{single}, results in a change in sign of the energy. The coherence length of the bound state is given by $\xi=v_F/\Delta$.

A similar zero energy bound state  localized on the right hand boundary can also be constructed by taking $\theta=-i\pi/2$ instead giving a state which decays as $e^{-\Delta |L-x|/v_F}$. Both bound states occur at the poles of the  boundary S-matrices, $K_L, ~K_R$ but importantly are not solutions of the quantization condition \eqref{BAEfree}. If instead we choose the boundary conditions \eqref{ABC1} and \eqref{ABC2} we have that $K_L=K_R=\mathbb{1}$. Evidently this has no poles and does not support bound states but nevertheless, the spectrum is also determined by \eqref{BAEfree}.

\section{Non-interacting model}
Before proceeding to the fully interacting model, it is instructive to construct the ground state and excitations of the free model which follows a similar methodology. The $N$-particle eigenstates when $g=0$ are merely products over the single particle states, \eqref{single} with rapidities  $\theta_j,~j=1,\dots,N$. The energy of this state is the sum over single particle energies, $\sum_j^N\epsilon(\theta_j)$. In the thermodynamic limit, $N,L\to\infty$ this sum can be changed to an integral $\sum_j^N\to L\int_0^\Lambda \mathrm{d}\theta\rho_0(\theta)$ where $\Lambda$ is a cutoff imposed on the rapidities and $\rho_0(\theta)$ is the distribution of rapidities in the state. This distribution is defined in the standard way as \cite{Takahashi, KorepinBook, Gaudin, 4lectures}
\begin{eqnarray}\label{rho}
\rho_0(\theta_j)=\frac{n_{j+1}-n_j}{L(\theta_{j+1}-\theta_j)}
\end{eqnarray}
where $n_j$ are the integer quantum numbers of the state. They arise from taking the logarithm of \eqref{BAEfree} such that $n_j=\Delta\sinh{(\theta_j)}L/v_F$. Note that $n_j$ may be both positive or negative depending on the imaginary part of $\theta_j$ but must all be distinct and  non-zero in order to have a non-vanishing wavefunction. 

The ground state is constructed by taking the values $n_j=-j$ so that it consists of negative energy particles with no holes. Using this along with \eqref{rho} we find that the ground state distribution in the thermodynamic limit is
\begin{eqnarray}\label{freerho}
\rho_0(\theta)=\frac{\Delta}{\pi v_F}\cosh{(\theta)}-\frac{\delta(\theta)}{L}.
\end{eqnarray}
Here we have subtracted off a delta function so that the distribution is defined for $\theta\geq 0$ and the hole at $\theta=0$ is accounted for.
The cutoff is then fixed by using $D=\int_0^\Lambda\mathrm{d}\theta \rho_0(\theta)$ where $D=N/L$ is the density. From this we have that $\Lambda=\log{(2\pi v_F D'/\Delta)}$ with $D'=D+1/L$. The ground state energy density is then simply given by \begin{eqnarray}
\varepsilon_0&=&-\int_0^\Lambda\mathrm{d}\theta\, \rho_0(\theta)\Delta\cosh{\theta}\\\label{gsenergy}
&=&-\frac{\pi \hbar v_F}{2}D'^2-\frac{\Delta^2}{2\pi}\log{\left(\frac{2\pi v_FD'}{\Delta}\right)}+\frac{\Delta}{L}.
\end{eqnarray}
In the non-interacting model the ground state (and all other states) enjoys a large degeneracy coming from the amplitude of the wavefunction which takes the form $\left[\otimes_{j=1}^M(0,1)^T\right]\otimes \left[\otimes_{j=1}^{N-M}(1,0)^T\right] $ for $M\leq N/2$. This degeneracy coming from the decoupling of the species is completely lifted in the interacting case. 

The excitations above this ground state consist of either removing negative energy particle, at say $\theta=\theta_h+i\pi$ from the ground state or adding a positive energy particle on top of this at $ \theta =\theta_p$. In the former case we modify ground state quantum numbers so that one of the integers, $n_j$ is missing. The distribution is then modified so as to include this hole $\rho_0(\theta)\to \rho_0(\theta)-\delta(\theta-\theta_h)/L$. For either, the energy increase is $\Delta \cosh{(\theta_{h,p})}$ and the particle number is changed  by $\delta N=\pm 1$. Particle-hole excitations, i.e. a state with both types of excitations,  leave the particle number unchanged and have a minimum energy of $2\Delta$ which is the energy gap of the non-interacting system.

On top of this state we may add either one or two bound states with no change in the energy meaning that the ground state has a further four-fold degeneracy. Unlike the typical degeneracy associated to a free model discussed above, the addition of bound states  changes the total fermion number to $N_\text{total}=N+n_\text{B}$ with $n_\text{B}$ the number of bound states in the system. Consequently the total fermion parity, defined as $\mathcal{P}=(-1)^{N+n_\text{B}}$ may be changed by their inclusion. 

In the interacting model we will see that by adding particles or holes to the ground state the distribution is shifted. This leads to a change in the dispersion relation of the excitations as well as the energy gap. The same is true when bound states are added to the system, the ground state distribution is shifted and their associated  degeneracy is lifted. 


\section{Many-body eigenstates}
In the interacting model the many-particle eigenstates can be constructed in the standard Bethe Ansatz fashion\cite{ 4lectures, KorepinBook}. The $N$-particle scattering eigenstate of energy $E=\sum_{j=1}^N\epsilon(\theta_j)$ is given by
\begin{eqnarray}\label{State}
\ket{\vec{\epsilon}}=\sum_{\vec{\xi},\vec{\sigma},Q}\int_0^L\!\mathrm{d}x\, A^{\vec{\xi}}_{\vec{\sigma}}[Q]\Theta(x_Q)\prod_{j=1}^N\Psi^\dag_{\xi_j}(x_j,\sigma_j\theta_j)\ket{0}~
\end{eqnarray}
where $\Theta(x_Q)$ are Heaviside functions which are non-zero only for a particular ordering of particles e.g, $x_1<x_2<\dots x_N$. The orderings of particles are labelled by $Q$ which are elements of the symmetric group of $N$ objects, $\mathcal{S}_N$.  We sum over all such orderings as well as combinations $\vec{\sigma}=(\sigma_1,\dots,\sigma_N)$ with $\sigma_j=\pm$ and $\vec{\xi}=(\xi_1,\dots,\xi_N)$ with $\xi_j=1,2$. The coefficients $A^{\vec{\xi}}_{\vec{\sigma}}[Q]$ are the amplitudes for one specific choice of $Q,\vec{\sigma}$ and $\vec{\xi}$ and are related to each other by products of the boundary S-matrices, $K_{L,R}$ (see above \eqref{K} for the analogous condition in the non-interacting case) and two particle S-matrices $S_{ij}$.
These two particle S-matrices act on the internal space of the $i^\text{th}$ and $j^\text{th}$ particle and are determined to be
\begin{eqnarray}
S_{ij}(\theta_i-\theta_j)=\begin{pmatrix}
1&0&0&0\\
0& \frac{\sinh{[\theta_i-\theta_j]}}{\sinh{[\theta_i-\theta_j-i\delta]}}&\frac{\sinh{[-i\delta]}}{\sinh{[\theta_i-\theta_j-i\delta]}} & 0\\
0&\frac{\sinh{[-i\delta]}}{\sinh{[\theta_i-\theta_j-i\delta]}} & \frac{\sinh{[\theta_i-\theta_j]}}{\sinh{[\theta_i-\theta_j-i\delta]}}& 0\\
0&0&0&1
\end{pmatrix}
\end{eqnarray}
where $\delta=2\arctan{(g/2v_F)}$ is the two particle phase shift. In deriving this relationship between the phase shift $\delta$ and the interaction strength we have chosen a specific regularization of the delta function interaction in the model. The functional form of $\delta(g)$ depends upon this and is therefore not universal except at small $g$ where $\delta\approx g/v_F$\footnote{A discussion of the different choices of regularization is presented in [\citenum{Roy}]. See [\citenum{BergknoffThacker}] and [\citenum{Korepinmtm}] for two different choices in the sine-Gordon model with different expressions for the phase shift in terms of the interaction strength. The regularization used in the text corresponds to that which is used in [\citenum{BergknoffThacker}].  Also see [\citenum{Camacho}] for a  discussion in other models and a comparison with bosonization} with  positive and negative $\delta$  corresponding to the repulsive and attractive regimes respectively. However the relationship between between $\delta$ and $K_\pm$ is universal and given by\cite{SaleurBL, RylandsAndrei}
\begin{eqnarray}
K_-=1-\frac{\delta}{\pi},~~
K_+=\frac{1}{1+\delta/\pi}
\end{eqnarray}
where we are restricted to $\delta\in[-\pi,\pi]$.

As in the non-interacting case the particle rapidities are quantized by applying the boundary conditions at $x=0,L$. This leads to an eigenvalue problem, similar to \eqref{BAEfree}, to determine the $\theta_j$
\begin{eqnarray}\label{Quant}
e^{-2i\Delta\sinh{(\theta_j)}L/v_F}A_{-}[\mathbb{1}]=Z_j(\theta_j)A_{-}[\mathbb{1}]
\end{eqnarray}
with 
\begin{eqnarray}\nonumber
Z_j(\theta_j)&=&S_{jj-1}(-\theta_j-\theta_{j-1})...S_{j1}(-\theta_j-\theta_1)K_R(\theta_j)\\\nonumber
&&\times S_{j1}(\theta_j-\theta_1)...S_{jN}(\theta_j-\theta_1)K_L(\theta_j)\\\label{Z}
&&\times S_{jN}(-\theta_j-\theta_N)... S_{jj+1}(-\theta_j-\theta_{j+1})
\end{eqnarray}
 and where $A_-[\mathbb{1}]$ is the amplitude for the  configuration $x_j<x_k$ for $j<k$ and $\sigma_j=-1, ~\forall j$. This can be interpreted in the same way as the non-interacting model. The eigenvalues of the object $Z_j(\theta_j)$ are the total   phase shift acquired the particle as it traverses the system, scattering off all other particles and both boundaries as it does so. In the non-interacting limit two particle S-matrices become identities and we recover $Z_j(\theta_j)=K_R(
 \theta_j)K_L(\theta_j)=\mathbb{1}$.
 
 Before discussing the solution of this eigenvalue equation we examine the case where bound states are present also.
In the same manner as \eqref{State} we may also construct many particle eigenstates which include either or both bound states at the edges of the system. Adding these on top of the previously constructed $N$-particle eigenstate we have 
\begin{eqnarray}
\ket{\vec{\epsilon}}_{\text{B}}=\sum'_{\vec{\xi},\vec{\sigma},Q}\int_0^L\!\mathrm{d}x\, A^{\vec{\xi}}_{\vec{\sigma}}[Q]\Theta(x_Q)\prod_{j=0}^{N+1}\Psi^\dag_{\xi_j}(x_j,\sigma_j\theta_j)\ket{0}~
\end{eqnarray}
where $x_0, x_{N+1}$ are the coordinates for the bound states at the left and right boundary and the sum over orderings now extends to elements of $\mathcal{S}_{N+2}$ as there are $N+2$ particles in total. In addition, the primed sum indicates that we sum over all possible flavor combinations but we restrict to $\sigma_0=\sigma_{N+1}=+$ and $\theta_0=-\theta_{N+1}=i\pi/2$. Note that the coherence length of the bound state remains unchanged in the presence of interactions $\xi=v_F/\Delta$ as is the case for Majorana bound states\cite{Gangadharaiah, Stoudenmire}.

The energy of this eigenstate, being the sum of single particle energies, is of the same form as that in which there are no bound states $E=\sum_{j=1}^{N}\epsilon(\theta_j)$.  However the $\theta_j$ are coupled together and their allowed values are shifted by the presence of the bound states. In particular when there are bound states both of the  edges we arrive the boundary conditions impose  the equations \eqref{Quant} and \eqref{Z} but with the boundary matrices replaced with $K_{L,R}\to K_{L,R}^\text{B}$ where
\begin{eqnarray}\nonumber
K^\text{B}_{R}(\theta_j)&=&S_{N+1}(-\theta_j+i\pi/2)K_R(\theta_j)S_{N+1}(\theta_j+i\pi/2),\\
K^\text{B}_{L}(\theta_j)&=&S_{0j}(\theta_j-i\pi/2)K_L(\theta_j)S_{0j}(-\theta_j-i\pi/2).
\end{eqnarray}
Alternatively one may consider a bound state at only one of the boundaries by replacing only one of  $K_{L,R}$.


\section{Bethe Ansatz equations}
The spectrum of our model can be determined by obtaining the eigenvalues of the operator $Z_j(\theta_j)$  $\forall j$. This can be solved by means of the off-diagonal Bethe Ansatz method and in fact it maps directly onto the solution of the inhomogeneous XXZ model with certain open boundary conditions \cite{ODBA}. To make use of this solution and simplify calculation somewhat we modify the right hand boundary so that it is given by $K'_R(\theta)=K_R(\theta-i\delta)$
 \footnote{This change of the right hand boundary condition results in a shift of the boundary contribution to the energy density, however the boundary energy comes from a redistribution of the Fermi sea which decays exponentially away from the Fermi level. At the Fermi level the modification amounts to an additional constant phase shift acquired by particle reflecting off the boundary which will not provide a shift in the boundary energy. As such in the thermodynamic limit the modification results in a negligible change in the energy. The bound state energy is unaffected by this shift. }.
This reduces to the previous case in the non-interacting limit and also generalizes the relationship between the two boundaries to $K_R(
\theta)=K_L(-\theta+i\delta)$ which is known as boundary crossing invariance \cite{Ghoshal}. With this modification we find that the particle rapidities $\theta_j$  are quantized according to 
\begin{eqnarray}\label{Bethe1}
e^{-2i\Delta \sinh{(\theta_j)}L/v_F}=\prod_{\sigma=\pm}\prod_{\alpha=1}^M\frac{\sinh{[\theta_j-\sigma\mu_\alpha+i\delta/2]}}{\sinh{[\theta_j-\sigma\mu_\alpha-i\delta/2]}}
\end{eqnarray}
where the parameters, $\mu_\alpha$ are known as Bethe parameters which describe the flavor degrees of freedom. They are determined by the following equations
\begin{eqnarray}\nonumber
\left[\frac{\cosh^2{[\mu_j+i\delta/2]}}{\cosh^2{[\mu_j-i\delta/2]}}\right]^{2-n_\text{B}}\prod_{\sigma=\pm}\prod_{\beta\neq \alpha}^M\frac{\sinh{[\mu_\alpha+\sigma\mu_\beta+i\delta]}}{\sinh{[\mu_\alpha+\sigma\mu_\beta-i\delta]}}\\\label{Bethe2}
=\prod_{\sigma=\pm}\prod_{k=1}^N\frac{\sinh{[\mu_\alpha+\sigma\theta_k+i\delta/2]}}{\sinh{[\mu_\alpha+\sigma\theta_k-i\delta/2]}}~~~~~
\end{eqnarray}
where, as before $M\leq N/2$ is an integer. We consider only positive  values of $\theta_j,\mu_\alpha$ and furthermore all rapidities and Bethe parameters must be distinct, $\theta_j\neq \theta_k$, $\mu_\alpha\neq \mu_\beta$  and non zero otherwise the corresponding wavefunction vanishes \cite{Sklyannin, KorepinBook}. The set of equations given in \eqref{Bethe1} and \eqref{Bethe2} are the Bethe Ansatz equations. 

The first term on the left hand side of \eqref{Bethe2} is the combined phase shift a particle accumulates after scattering off both boundaries and any bound states which are attached to them. If we were to consider the alternative boundary  condition given by \eqref{ABC1} and \eqref{ABC2}  which does not mix the flavors this term would be absent. Interestingly this has the same effect on the Bethe parameters and therefore spectrum as the presence of both bound states i.e. $n_\text{B}=2$.  The next term on the left hand side is due to the interaction with other particles within the bulk of the system. These terms vanish upon taking $\delta=0$ whereupon we recover \eqref{BAEfree}. 

The Bethe equations also reduce to \eqref{BAEfree} when $\delta=\pm\pi$ indicating that there is a self-duality in the theory, i.e. a mapping from the model to itself at a different value of the interaction strength\cite{FendleySaleur}. To investigate this further, note that the Bethe equations are invariant under the combined transformation $\delta\to-\delta$ along with $\theta_j\to \theta_j+i\pi$ however this changes the sign of the energy, $\sum_j\epsilon(\theta_j)\to -\sum_j\epsilon(\theta_j)$. Thus the spectrum is inverted under a change in sign of the interaction strength. This is a manifestation of the particle-hole transformation of our original model. Furthermore it can be checked that the  replacement $\delta\to \pi-|\delta|$ along with a redefinition $\mu_\alpha\to\mu+i\pi/2$ has the same effect on the Bethe equations as taking $\delta\to -\delta$. Combining these two maps therefore leaves the spectrum invariant and allows us to restrict our analysis to $\delta\in[0,\pi/2]$ with  results outside this region found using the above transformations.

\section{Ground state}
The structure of the Bethe equations is similar to those appearing in the solutions of a number of other models\cite{BL, SaleurBL,RylandsAndrei} and the present analysis follows similar lines using the methodology presented for the non-interacting case.

When $0\leq\delta\leq \frac{\pi}{2}$ the ground state consists of all $\theta_j$ lying on the $i\pi$ line and $M=N/2$. Using this in \eqref{Bethe1} and \eqref{Bethe2} and then taking their logarithm we have 
\begin{eqnarray}\label{logBAE}
\frac{\Delta}{\pi v_F}\sinh{(\theta_j)}L&=& n_j-\sum_{\substack{\alpha\\\sigma=\pm}}^{N/2}\phi_1(\theta_j-\sigma\mu_\alpha)\\\nonumber
\sum_{\substack{j=1\\\sigma=\pm}}^{N}\phi_1(\mu_\alpha-\sigma\theta_j)&=& I_j+\sum_{\substack{\beta=1\\\sigma=\pm}}^{N/2}\phi_2(\mu_\alpha-\sigma\mu_\beta)+g(\mu_\alpha)
\end{eqnarray}
Here $n_j$ and $I_j$ are integers which are the quantum numbers of the interacting system and $\phi_n(x,y)=\frac{i}{2\pi}\log{[\frac{\sinh{(x+in\delta/2)}}{\sinh{(x-in\delta/2)}}]}$, ~$g(\mu)=i\frac{2-n_\text{B}}{\pi}\log[\frac{\cosh{[\mu_j+i\delta/2]}}{\cosh{[\mu_j-i\delta/2]}}]$. In the thermodynamic limit we may describe the ground state via the rapidity distribution $\rho(\theta)$, defined by \eqref{rho} and the analogous distribution for the Bethe parameters $\nu(\mu)$ which is defined similarly. In the thermodynamic limit, \eqref{logBAE} become a set of coupled integral equations,
\begin{eqnarray}\nonumber
\frac{\Delta}{\pi v_F}\cosh{(\theta)}-\frac{\delta(\theta)}{L}&=&\rho(\theta)-\int\mathrm{d}\mu\, \phi_1'(\theta-\mu)\nu(\mu)~\\\nonumber
\int\mathrm{d}\theta \,\phi_1' (\mu-\theta)\rho(\theta)-\frac{\delta(\mu)}{L}&=&\nu(\mu)+\int\mathrm{d}\zeta\, \phi_2'(\mu-\zeta) \nu(\zeta)\\
&&-\frac{\phi_2'(2\mu)}{L}-\frac{g'(\mu)}{L}
\end{eqnarray}
where the delta functions are included to account for the holes at $\theta=\mu=0$. These equations may be solved via Fourier transform with the result
\begin{eqnarray}\label{rhobulk}
\rho(\theta)&=&\frac{2\Delta}{\pi v_F}\cosh{(\theta)}+\rho_\text{bdry}(\theta)+\rho_\text{B}(\theta)\\
\nu(\mu)&=&\frac{\Delta \cosh{(\mu)}}{\pi v_F \cos{(\delta/2)}}+\nu_\text{bdry}(\mu)+\nu_\text{B}(\mu).
\end{eqnarray}
The first terms in the above expressions correspond to the bulk contribution, note that the rapidity distribution is modified in the interacting case as compared to \eqref{freerho}. The next terms arise due to the presence of the boundary, they are distinguished from the bulk by being of order $1/L$ and are  independent of the number of bound states. The last terms are those which are attributable to the bound states and are proportional to $n_\text{B}/L$. The total energy density of the state is determined solely by the rapidity distribution via $\varepsilon_g=-\int\mathrm{d}\theta\,\rho(\theta)\Delta\cosh{(\theta)}$ giving
\begin{eqnarray}\nonumber
\varepsilon_g=\varepsilon_0+\frac{\Delta}{L}\left[\frac{1}{2\cos{(\delta/2)}}-\sqrt{2}\cos{(\delta/2)}\right]+\frac{n_\text{B}-2}{L}\epsilon_\text{B}.
\end{eqnarray}
The first term here is the bulk energy density which is given by \eqref{gsenergy} modulo a $\delta$ dependent shift in the $D'$ \footnote{In order to compare the energy of the states with and without the bound states present the cutoff must be the same for both. The cutoff for the interacting case, $\Lambda_g$, is determined using $D=\int_0^{\Lambda_g}\mathrm{d}\theta\,\rho(\theta)$ in the absence of the bound states. It is then held fixed and the energy with and without the bound states is calculated and compared. Alternatively we could have determined the cutoff with the bound states included already with the same result. } which vanishes in the thermodynamic limit, the remaining terms are due to the boundary conditions and bound states with $\epsilon_\text{B}$ being the energy per bound state.

Before discussing this bound state contribution to the energy we shall comment on the excitations of the model. The lowest lying excitations come in two forms, the first of which are similar to the non-interacting case. They can be created by   placing hole in the rapidity distribution at $\theta=\theta_h+i\pi$  or adding a positive energy particle at $\theta=\theta_p$,  in either case the energy is given by $\epsilon_1(\theta_{h,p})=2\Delta\cosh{(\theta_{h,p})}$. The factor of two present here as compared to the non-interacting case can be traced to the overall factor of two in \eqref{rhobulk}. A particle-hole excitation of this type has a minimum energy of $4\Delta$ which is twice that of the non interacting model. The second type of excitation involves placing a hole in the distribution of Bethe roots at say $\mu=\theta_h$. In this case the energy is given by $\epsilon_2(\theta_h)=\frac{\Delta \cosh{(\theta_h)}}{\cos{(\delta/2)}}$. As a result the gap in the interacting system is increased to  $2\Delta_g$ where 
\begin{eqnarray}
\Delta_g=\frac{\Delta}{\cos{(\delta/2)}}.
\end{eqnarray}
The bound state contribution to the rapidity distribution is used to determine the bound state energy via $\epsilon_\text{B}=-L\int\mathrm{d}\theta \rho_\text{B}(\theta)\Delta\cosh{(\theta)}/n_\text{B}$ where
\begin{eqnarray}\nonumber
\rho_\text{B}(\theta)=-\frac{n_\text{B}}{L}\int\frac{\mathrm{d}\omega}{2\pi}\frac{e^{-i\omega \theta}\sinh{[\pi\omega/2]}\sinh{[\delta \omega/2]}}{\sinh{[(\pi-\delta)\omega/2]}\sinh{[(\pi+\delta)\omega/2]}}
\end{eqnarray}
from which we have that the energy per bound state for $0\leq\delta\leq \pi/2$ is given by $\varepsilon_\text{B}=\Delta_g\tan{[\delta/2]}=\Delta_g\sqrt{(\Delta_g/\Delta)^2-1}.$ We can then combine this with the various symmetries of the Bethe equations discussed above and reconstruct the bound state energy for all values of $\delta$ to be
\begin{eqnarray}\label{Boundstate}
\varepsilon_\text{B}=\begin{cases}
\Delta_g\tan{(\delta/2)} &0\leq|\delta|\leq \frac{\pi}{2}\\
\Delta_g\cot{(\delta/2)}& \frac{\pi}{2}\leq|\delta|\leq \pi
\end{cases}
\end{eqnarray}
We see that the bound state energy is pushed below the Fermi level for $
\delta<0$ and above it for $\delta>0$ indicating that there is a first order phase transition at $\delta=0$. For attractive interactions the ground state consists of a filled Fermi sea with bound states at both ends.  Similar transitions are known to occur when magnetic impurities are present in superconductors\cite{KondoSCRMP, KondoPRX, Sakurai,SakaiShimizu, Yu, Rusinov, Shiba, Zitt, Iucci}. Therein, Shiba or Andreev bound states form at the impurity. In such cases the transition is accompanied by a change in the fermion parity of the ground state. In the present case this does not occur as bound states will be present at both ends of the wire, leaving $\mathcal{P}$ invariant.

For $|\delta|\leq \pi/2$, $|\varepsilon_\text{B}|$ increases and at $\delta=\pm\pi/2$ it touches the continuum of states in the conduction or valence bands. Upon further increasing $|\delta|$, $\varepsilon_\text{B}$ undergoes a sharp change in behaviour, turning away from the bands and again approaching the Fermi level.  Recall that there are holes at $\theta=0$ and so the bound state does not become degenerate with any scattering state. This sharp avoided level crossing is a consequence of the self-duality of the model, i.e. the invariance of the spectrum under the combined transformations, $\delta\to-\delta$ and then $\delta\to\pi-|\delta|$, with the avoided crossing occurring at the special point $\delta=\pm\pi/2$ which corresponds to $\xi\Delta_g/v_F=\sqrt{2}$. As a result the bound states cannot couple to scattering states and are protected from decay. Therefore in contrast to non-interacting systems where boundary modes are protected by symmetry, in the interacting model they are protected by self-duality.

 The bound state energy reaches the Fermi level once again at the strongly interacting points $\delta=\pm \pi$ indicating that the symmetry of non-interacting model is restored. This can be seen also in the bosonic language also. At the points, $\delta=\pi$ the spin Luttinger parameter vanishes, $K_-=0$ while at $\delta=-\pi$ we have $1/K_+=0$. In either case  the fields $\phi_\pm$ decouple from each other with only one being gapped. In these cases the bosonic model, \eqref{Hbosdual} can be mapped to a quadratic model of spinless fermions\cite{Gangadharaiah}. Therein the bound states  lie at the Fermi level  in agreement with \eqref{Boundstate} derived using self duality.

\section{Conclusions}

In this paper we have studied the boundary bound states of a one dimensional, spinful quantum wire. The wire Hamiltonian has a proximity induced $p_x$-wave triplet pairing and density-density interactions. 
We have solved this system exactly using Bethe Ansatz for a range of parameters and constructed the ground state and excitations of the model. It was shown that 
for a choice of boundary conditions which breaks time reversal symmetry, the system can support bound states localized at both ends. The bound state energy lies at the Fermi level, within the energy gap when interactions are absent but is shifted when interactions are present. For attractive interactions the bound state energy is pushed below the Fermi level while in the repulsive case it is pushed up. This causes a first order phase transition to occur at zero temperature. Unlike similar transitions in superconductor-impurity systems this is not accompanied by a change in fermion parity. 

The bound state energy is seen to approach the gap, $\Delta_g$ as the interaction strength, $|\delta|$ is increased but undergoes a sharp avoided level crossing at $\delta=\pm\pi/2$ thus preventing the bound state from entering the continuum of scattering states and decaying. This is a consequence of the self-duality of the model which relates the spectrum of model at different values of the interaction strength.

In the non-interacting case the bound states lead to a Lorentzian, zero-bias peak in the conductance through the edge \cite{Sengupta}. When interactions are present this peak will be shifted owing to the non-zero energy of the bound states but shall remain in the gap. In addition the Lorentzian shape of the peak is maintained in the presence of interactions owing to the fact that pairing term has scaling dimension 1 irrespective of the value of $\delta$\cite{BL, SaleurBL}. For a more general form of the interaction, which may break the integrability of the model, one can expect that the scaling dimension becomes a function of $\delta$ leading to an energy gap which has a power law dependence on $\Delta$ as well as a power law decay of the conductance away from the peak\cite{Giamarchi, GogolinNerseyanTsvelik, Tsvelik}. 


\acknowledgements
 Helpful discussions with Natan Andrei, Daniel Brennan, Victor Galitski and Anibal Iucci are gratefully acknowledged.
 
 
\bibliography{bib}

\providecommand{\noopsort}[1]{}\providecommand{\singleletter}[1]{#1}%
\begin{thebibliography}{61}%
\makeatletter
\providecommand \@ifxundefined [1]{%
 \@ifx{#1\undefined}
}%
\providecommand \@ifnum [1]{%
 \ifnum #1\expandafter \@firstoftwo
 \else \expandafter \@secondoftwo
 \fi
}%
\providecommand \@ifx [1]{%
 \ifx #1\expandafter \@firstoftwo
 \else \expandafter \@secondoftwo
 \fi
}%
\providecommand \natexlab [1]{#1}%
\providecommand \enquote  [1]{``#1''}%
\providecommand \bibnamefont  [1]{#1}%
\providecommand \bibfnamefont [1]{#1}%
\providecommand \citenamefont [1]{#1}%
\providecommand \href@noop [0]{\@secondoftwo}%
\providecommand \href [0]{\begingroup \@sanitize@url \@href}%
\providecommand \@href[1]{\@@startlink{#1}\@@href}%
\providecommand \@@href[1]{\endgroup#1\@@endlink}%
\providecommand \@sanitize@url [0]{\catcode `\\12\catcode `\$12\catcode
  `\&12\catcode `\#12\catcode `\^12\catcode `\_12\catcode `\%12\relax}%
\providecommand \@@startlink[1]{}%
\providecommand \@@endlink[0]{}%
\providecommand \url  [0]{\begingroup\@sanitize@url \@url }%
\providecommand \@url [1]{\endgroup\@href {#1}{\urlprefix }}%
\providecommand \urlprefix  [0]{URL }%
\providecommand \Eprint [0]{\href }%
\providecommand \doibase [0]{http://dx.doi.org/}%
\providecommand \selectlanguage [0]{\@gobble}%
\providecommand \bibinfo  [0]{\@secondoftwo}%
\providecommand \bibfield  [0]{\@secondoftwo}%
\providecommand \translation [1]{[#1]}%
\providecommand \BibitemOpen [0]{}%
\providecommand \bibitemStop [0]{}%
\providecommand \bibitemNoStop [0]{.\EOS\space}%
\providecommand \EOS [0]{\spacefactor3000\relax}%
\providecommand \BibitemShut  [1]{\csname bibitem#1\endcsname}%
\let\auto@bib@innerbib\@empty
\bibitem [{\citenamefont {Di~Francesco}\ \emph {et~al.}(1996)\citenamefont
  {Di~Francesco}, \citenamefont {Mathieu},\ and\ \citenamefont
  {S{\'e}n{\'e}chal}}]{DiFrancesco}%
  \BibitemOpen
  \bibfield  {author} {\bibinfo {author} {\bibfnamefont {P.}~\bibnamefont
  {Di~Francesco}}, \bibinfo {author} {\bibfnamefont {P.}~\bibnamefont
  {Mathieu}}, \ and\ \bibinfo {author} {\bibfnamefont {D.}~\bibnamefont
  {S{\'e}n{\'e}chal}},\ }\href {https://books.google.ca/books?id=mcMbswEACAAJ}
  {\emph {\bibinfo {title} {Conformal Field Theory}}},\ Graduate texts in
  contemporary physics\ (\bibinfo  {publisher} {Island Press},\ \bibinfo {year}
  {1996})\BibitemShut {NoStop}%
\bibitem [{\citenamefont {Sachdev}(2011)}]{Sachdev}%
  \BibitemOpen
  \bibfield  {author} {\bibinfo {author} {\bibfnamefont {Subir}\ \bibnamefont
  {Sachdev}},\ }\href {\doibase 10.1017/CBO9780511973765} {\emph {\bibinfo
  {title} {Quantum Phase Transitions}}},\ \bibinfo {edition} {2nd}\ ed.\
  (\bibinfo  {publisher} {Cambridge University Press},\ \bibinfo {year}
  {2011})\BibitemShut {NoStop}%
\bibitem [{\citenamefont {Halperin}(1982)}]{Halperin}%
  \BibitemOpen
  \bibfield  {author} {\bibinfo {author} {\bibfnamefont {B.~I.}\ \bibnamefont
  {Halperin}},\ }\bibfield  {title} {\enquote {\bibinfo {title} {Quantized hall
  conductance, current-carrying edge states, and the existence of extended
  states in a two-dimensional disordered potential},}\ }\href {\doibase
  10.1103/PhysRevB.25.2185} {\bibfield  {journal} {\bibinfo  {journal} {Phys.
  Rev. B}\ }\textbf {\bibinfo {volume} {25}},\ \bibinfo {pages} {2185--2190}
  (\bibinfo {year} {1982})}\BibitemShut {NoStop}%
\bibitem [{\citenamefont {Hasan}\ and\ \citenamefont {Kane}(2010)}]{Hasan}%
  \BibitemOpen
  \bibfield  {author} {\bibinfo {author} {\bibfnamefont {M.~Z.}\ \bibnamefont
  {Hasan}}\ and\ \bibinfo {author} {\bibfnamefont {C.~L.}\ \bibnamefont
  {Kane}},\ }\bibfield  {title} {\enquote {\bibinfo {title} {Colloquium:
  Topological insulators},}\ }\href {\doibase 10.1103/RevModPhys.82.3045}
  {\bibfield  {journal} {\bibinfo  {journal} {Rev. Mod. Phys.}\ }\textbf
  {\bibinfo {volume} {82}},\ \bibinfo {pages} {3045--3067} (\bibinfo {year}
  {2010})}\BibitemShut {NoStop}%
\bibitem [{\citenamefont {Qi}\ and\ \citenamefont {Zhang}(2011)}]{Qi}%
  \BibitemOpen
  \bibfield  {author} {\bibinfo {author} {\bibfnamefont {Xiao-Liang}\
  \bibnamefont {Qi}}\ and\ \bibinfo {author} {\bibfnamefont {Shou-Cheng}\
  \bibnamefont {Zhang}},\ }\bibfield  {title} {\enquote {\bibinfo {title}
  {Topological insulators and superconductors},}\ }\href {\doibase
  10.1103/RevModPhys.83.1057} {\bibfield  {journal} {\bibinfo  {journal} {Rev.
  Mod. Phys.}\ }\textbf {\bibinfo {volume} {83}},\ \bibinfo {pages}
  {1057--1110} (\bibinfo {year} {2011})}\BibitemShut {NoStop}%
\bibitem [{\citenamefont {Kitaev}(2001)}]{Kitaev}%
  \BibitemOpen
  \bibfield  {author} {\bibinfo {author} {\bibfnamefont {A~Yu}\ \bibnamefont
  {Kitaev}},\ }\bibfield  {title} {\enquote {\bibinfo {title} {Unpaired
  majorana fermions in quantum wires},}\ }\href {\doibase
  10.1070/1063-7869/44/10s/s29} {\bibfield  {journal} {\bibinfo  {journal}
  {Physics-Uspekhi}\ }\textbf {\bibinfo {volume} {44}},\ \bibinfo {pages}
  {131–136} (\bibinfo {year} {2001})}\BibitemShut {NoStop}%
\bibitem [{\citenamefont {Lutchyn}\ \emph {et~al.}(2010)\citenamefont
  {Lutchyn}, \citenamefont {Sau},\ and\ \citenamefont {Das~Sarma}}]{Lutchyn}%
  \BibitemOpen
  \bibfield  {author} {\bibinfo {author} {\bibfnamefont {Roman~M.}\
  \bibnamefont {Lutchyn}}, \bibinfo {author} {\bibfnamefont {Jay~D.}\
  \bibnamefont {Sau}}, \ and\ \bibinfo {author} {\bibfnamefont
  {S.}~\bibnamefont {Das~Sarma}},\ }\bibfield  {title} {\enquote {\bibinfo
  {title} {Majorana fermions and a topological phase transition in
  semiconductor-superconductor heterostructures},}\ }\href {\doibase
  10.1103/physrevlett.105.077001} {\bibfield  {journal} {\bibinfo  {journal}
  {Physical Review Letters}\ }\textbf {\bibinfo {volume} {105}} (\bibinfo
  {year} {2010}),\ 10.1103/physrevlett.105.077001}\BibitemShut {NoStop}%
\bibitem [{\citenamefont {Oreg}\ \emph {et~al.}(2010)\citenamefont {Oreg},
  \citenamefont {Refael},\ and\ \citenamefont {von Oppen}}]{Oreg}%
  \BibitemOpen
  \bibfield  {author} {\bibinfo {author} {\bibfnamefont {Yuval}\ \bibnamefont
  {Oreg}}, \bibinfo {author} {\bibfnamefont {Gil}\ \bibnamefont {Refael}}, \
  and\ \bibinfo {author} {\bibfnamefont {Felix}\ \bibnamefont {von Oppen}},\
  }\bibfield  {title} {\enquote {\bibinfo {title} {Helical liquids and majorana
  bound states in quantum wires},}\ }\href {\doibase
  10.1103/physrevlett.105.177002} {\bibfield  {journal} {\bibinfo  {journal}
  {Physical Review Letters}\ }\textbf {\bibinfo {volume} {105}} (\bibinfo
  {year} {2010}),\ 10.1103/physrevlett.105.177002}\BibitemShut {NoStop}%
\bibitem [{\citenamefont {Shiba}(1968)}]{Shiba}%
  \BibitemOpen
  \bibfield  {author} {\bibinfo {author} {\bibfnamefont {Hiroyuki}\
  \bibnamefont {Shiba}},\ }\bibfield  {title} {\enquote {\bibinfo {title}
  {{Classical Spins in Superconductors}},}\ }\href {\doibase
  10.1143/PTP.40.435} {\bibfield  {journal} {\bibinfo  {journal} {Progress of
  Theoretical Physics}\ }\textbf {\bibinfo {volume} {40}},\ \bibinfo {pages}
  {435--451} (\bibinfo {year} {1968})},\ \Eprint
  {http://arxiv.org/abs/http://oup.prod.sis.lan/ptp/article-pdf/40/3/435/5185550/40-3-435.pdf}
  {http://oup.prod.sis.lan/ptp/article-pdf/40/3/435/5185550/40-3-435.pdf}
  \BibitemShut {NoStop}%
\bibitem [{\citenamefont {{Rusinov}}(1969)}]{Rusinov}%
  \BibitemOpen
  \bibfield  {author} {\bibinfo {author} {\bibfnamefont {A.~I.}\ \bibnamefont
  {{Rusinov}}},\ }\bibfield  {title} {\enquote {\bibinfo {title}
  {{Superconductivity near a Paramagnetic Impurity}},}\ }\href@noop {}
  {\bibfield  {journal} {\bibinfo  {journal} {Soviet Journal of Experimental
  and Theoretical Physics Letters}\ }\textbf {\bibinfo {volume} {9}},\ \bibinfo
  {pages} {85} (\bibinfo {year} {1969})}\BibitemShut {NoStop}%
\bibitem [{\citenamefont {Yu}(1965)}]{Yu}%
  \BibitemOpen
  \bibfield  {author} {\bibinfo {author} {\bibfnamefont {Luh}\ \bibnamefont
  {Yu}},\ }\bibfield  {title} {\enquote {\bibinfo {title} {Bound state in
  superconductors with paramagnetic impurities},}\ }\href {\doibase
  10.7498/aps.21.75} {\bibfield  {journal} {\bibinfo  {journal} {Acta Physica
  Sinica}\ }\textbf {\bibinfo {volume} {21}},\ \bibinfo {eid} {75} (\bibinfo
  {year} {1965})}\BibitemShut {NoStop}%
\bibitem [{\citenamefont {Balatsky}\ \emph {et~al.}(2006)\citenamefont
  {Balatsky}, \citenamefont {Vekhter},\ and\ \citenamefont {Zhu}}]{KondoSCRMP}%
  \BibitemOpen
  \bibfield  {author} {\bibinfo {author} {\bibfnamefont {A.~V.}\ \bibnamefont
  {Balatsky}}, \bibinfo {author} {\bibfnamefont {I.}~\bibnamefont {Vekhter}}, \
  and\ \bibinfo {author} {\bibfnamefont {Jian-Xin}\ \bibnamefont {Zhu}},\
  }\bibfield  {title} {\enquote {\bibinfo {title} {Impurity-induced states in
  conventional and unconventional superconductors},}\ }\href {\doibase
  10.1103/RevModPhys.78.373} {\bibfield  {journal} {\bibinfo  {journal} {Rev.
  Mod. Phys.}\ }\textbf {\bibinfo {volume} {78}},\ \bibinfo {pages} {373--433}
  (\bibinfo {year} {2006})}\BibitemShut {NoStop}%
\bibitem [{\citenamefont {Jackiw}\ and\ \citenamefont
  {Rebbi}(1976)}]{JackiwRebbi}%
  \BibitemOpen
  \bibfield  {author} {\bibinfo {author} {\bibfnamefont {R.}~\bibnamefont
  {Jackiw}}\ and\ \bibinfo {author} {\bibfnamefont {C.}~\bibnamefont {Rebbi}},\
  }\bibfield  {title} {\enquote {\bibinfo {title} {Solitons with fermion number
  1/2},}\ }\href {\doibase 10.1103/PhysRevD.13.3398} {\bibfield  {journal}
  {\bibinfo  {journal} {Phys. Rev. D}\ }\textbf {\bibinfo {volume} {13}},\
  \bibinfo {pages} {3398--3409} (\bibinfo {year} {1976})}\BibitemShut {NoStop}%
\bibitem [{\citenamefont {Su}\ \emph {et~al.}(1979)\citenamefont {Su},
  \citenamefont {Schrieffer},\ and\ \citenamefont {Heeger}}]{SuShreiferHeeger}%
  \BibitemOpen
  \bibfield  {author} {\bibinfo {author} {\bibfnamefont {W.~P.}\ \bibnamefont
  {Su}}, \bibinfo {author} {\bibfnamefont {J.~R.}\ \bibnamefont {Schrieffer}},
  \ and\ \bibinfo {author} {\bibfnamefont {A.~J.}\ \bibnamefont {Heeger}},\
  }\bibfield  {title} {\enquote {\bibinfo {title} {Solitons in
  polyacetylene},}\ }\href {\doibase 10.1103/PhysRevLett.42.1698} {\bibfield
  {journal} {\bibinfo  {journal} {Phys. Rev. Lett.}\ }\textbf {\bibinfo
  {volume} {42}},\ \bibinfo {pages} {1698--1701} (\bibinfo {year}
  {1979})}\BibitemShut {NoStop}%
\bibitem [{\citenamefont {Takayama}\ \emph {et~al.}(1980)\citenamefont
  {Takayama}, \citenamefont {Lin-Liu},\ and\ \citenamefont
  {Maki}}]{TakaLinMaki}%
  \BibitemOpen
  \bibfield  {author} {\bibinfo {author} {\bibfnamefont {Hajime}\ \bibnamefont
  {Takayama}}, \bibinfo {author} {\bibfnamefont {Y.~R.}\ \bibnamefont
  {Lin-Liu}}, \ and\ \bibinfo {author} {\bibfnamefont {Kazumi}\ \bibnamefont
  {Maki}},\ }\bibfield  {title} {\enquote {\bibinfo {title} {Continuum model
  for solitons in polyacetylene},}\ }\href {\doibase 10.1103/PhysRevB.21.2388}
  {\bibfield  {journal} {\bibinfo  {journal} {Phys. Rev. B}\ }\textbf {\bibinfo
  {volume} {21}},\ \bibinfo {pages} {2388--2393} (\bibinfo {year}
  {1980})}\BibitemShut {NoStop}%
\bibitem [{\citenamefont {Wang}(1997)}]{Wang}%
  \BibitemOpen
  \bibfield  {author} {\bibinfo {author} {\bibfnamefont {Yupeng}\ \bibnamefont
  {Wang}},\ }\bibfield  {title} {\enquote {\bibinfo {title} {Exact solution of
  the open heisenberg chain with two impurities},}\ }\href {\doibase
  10.1103/PhysRevB.56.14045} {\bibfield  {journal} {\bibinfo  {journal} {Phys.
  Rev. B}\ }\textbf {\bibinfo {volume} {56}},\ \bibinfo {pages} {14045--14049}
  (\bibinfo {year} {1997})}\BibitemShut {NoStop}%
\bibitem [{\citenamefont {Skorik}\ and\ \citenamefont {Saleur}(1995)}]{Saleur}%
  \BibitemOpen
  \bibfield  {author} {\bibinfo {author} {\bibfnamefont {S.}~\bibnamefont
  {Skorik}}\ and\ \bibinfo {author} {\bibfnamefont {Hubert}\ \bibnamefont
  {Saleur}},\ }\bibfield  {title} {\enquote {\bibinfo {title} {Boundary bound
  states and boundary bootstrap in the sine-gordon model with dirichlet
  boundary conditions},}\ }\href {\doibase 10.1088/0305-4470/28/23/014}
  {\bibfield  {journal} {\bibinfo  {journal} {Journal of Physics A General
  Physics}\ }\textbf {\bibinfo {volume} {28}} (\bibinfo {year} {1995}),\
  10.1088/0305-4470/28/23/014}\BibitemShut {NoStop}%
\bibitem [{\citenamefont {Skorik}\ and\ \citenamefont
  {Kapustin}(1995)}]{Skorik}%
  \BibitemOpen
  \bibfield  {author} {\bibinfo {author} {\bibfnamefont {S.}~\bibnamefont
  {Skorik}}\ and\ \bibinfo {author} {\bibfnamefont {Anton}\ \bibnamefont
  {Kapustin}},\ }\bibfield  {title} {\enquote {\bibinfo {title} {Surface
  excitations and surface energy of the antiferromagnetic xxz chain by the
  bethe ansatz approach},}\ }\href {\doibase 10.1088/0305-4470/29/8/011}
  {\bibfield  {journal} {\bibinfo  {journal} {Journal of Physics A General
  Physics}\ }\textbf {\bibinfo {volume} {29}} (\bibinfo {year} {1995}),\
  10.1088/0305-4470/29/8/011}\BibitemShut {NoStop}%
\bibitem [{\citenamefont {LeClair}\ \emph {et~al.}(1995)\citenamefont
  {LeClair}, \citenamefont {Mussardo}, \citenamefont {Saleur},\ and\
  \citenamefont {Skorik}}]{LeClair}%
  \BibitemOpen
  \bibfield  {author} {\bibinfo {author} {\bibfnamefont {A.}~\bibnamefont
  {LeClair}}, \bibinfo {author} {\bibfnamefont {G.}~\bibnamefont {Mussardo}},
  \bibinfo {author} {\bibfnamefont {H.}~\bibnamefont {Saleur}}, \ and\ \bibinfo
  {author} {\bibfnamefont {S.}~\bibnamefont {Skorik}},\ }\bibfield  {title}
  {\enquote {\bibinfo {title} {Boundary energy and boundary states in
  integrable quantum field theories},}\ }\href@noop {} {\bibfield  {journal}
  {\bibinfo  {journal} {Nuclear Physics B}\ }\textbf {\bibinfo {volume}
  {453}},\ \bibinfo {pages} {581–618} (\bibinfo {year} {1995})}\BibitemShut
  {NoStop}%
\bibitem [{\citenamefont {Ghoshal}\ and\ \citenamefont
  {Zamolodchikov}(1994)}]{Ghoshal}%
  \BibitemOpen
  \bibfield  {author} {\bibinfo {author} {\bibfnamefont {Subir}\ \bibnamefont
  {Ghoshal}}\ and\ \bibinfo {author} {\bibfnamefont {Alexander}\ \bibnamefont
  {Zamolodchikov}},\ }\bibfield  {title} {\enquote {\bibinfo {title} {Boundary
  s matrix and boundary state in two-dimensional integrable quantum field
  theory},}\ }\href {\doibase 10.1142/s0217751x94001552} {\bibfield  {journal}
  {\bibinfo  {journal} {International Journal of Modern Physics A}\ }\textbf
  {\bibinfo {volume} {09}},\ \bibinfo {pages} {3841–3885} (\bibinfo {year}
  {1994})}\BibitemShut {NoStop}%
\bibitem [{\citenamefont {Grisaru}\ \emph {et~al.}(1995)\citenamefont
  {Grisaru}, \citenamefont {Mezincescu},\ and\ \citenamefont
  {Nepomechie}}]{GrisaruMezincescNepomechie}%
  \BibitemOpen
  \bibfield  {author} {\bibinfo {author} {\bibfnamefont {M~T}\ \bibnamefont
  {Grisaru}}, \bibinfo {author} {\bibfnamefont {L}~\bibnamefont {Mezincescu}},
  \ and\ \bibinfo {author} {\bibfnamefont {R~I}\ \bibnamefont {Nepomechie}},\
  }\bibfield  {title} {\enquote {\bibinfo {title} {Direct calculation of the
  boundary s-matrix for the open heisenberg chain},}\ }\href {\doibase
  10.1088/0305-4470/28/4/025} {\bibfield  {journal} {\bibinfo  {journal}
  {Journal of Physics A: Mathematical and General}\ }\textbf {\bibinfo {volume}
  {28}},\ \bibinfo {pages} {1027--1045} (\bibinfo {year} {1995})}\BibitemShut
  {NoStop}%
\bibitem [{\citenamefont {Wang}\ \emph {et~al.}(2015)\citenamefont {Wang},
  \citenamefont {Yang}, \citenamefont {Cao},\ and\ \citenamefont {Shi}}]{ODBA}%
  \BibitemOpen
  \bibfield  {author} {\bibinfo {author} {\bibfnamefont {Yupeng}\ \bibnamefont
  {Wang}}, \bibinfo {author} {\bibfnamefont {Wen-Li}\ \bibnamefont {Yang}},
  \bibinfo {author} {\bibfnamefont {Junpeng}\ \bibnamefont {Cao}}, \ and\
  \bibinfo {author} {\bibfnamefont {Kangjie}\ \bibnamefont {Shi}},\ }\href@noop
  {} {\emph {\bibinfo {title} {{Off-diagonal Bethe ansatz for exactly solvable
  models}}}}\ (\bibinfo  {publisher} {Springer},\ \bibinfo {address} {Berlin},\
  \bibinfo {year} {2015})\BibitemShut {NoStop}%
\bibitem [{\citenamefont {Grijalva}\ \emph {et~al.}(2019)\citenamefont
  {Grijalva}, \citenamefont {De~Nardis},\ and\ \citenamefont
  {Terras}}]{Grijalva}%
  \BibitemOpen
  \bibfield  {author} {\bibinfo {author} {\bibfnamefont {Sebastian}\
  \bibnamefont {Grijalva}}, \bibinfo {author} {\bibfnamefont {Jacopo}\
  \bibnamefont {De~Nardis}}, \ and\ \bibinfo {author} {\bibfnamefont
  {Véronique}\ \bibnamefont {Terras}},\ }\bibfield  {title} {\enquote
  {\bibinfo {title} {Open xxz chain and boundary modes at zero temperature},}\
  }\href {\doibase 10.21468/scipostphys.7.2.023} {\bibfield  {journal}
  {\bibinfo  {journal} {SciPost Physics}\ }\textbf {\bibinfo {volume} {7}}
  (\bibinfo {year} {2019}),\ 10.21468/scipostphys.7.2.023}\BibitemShut
  {NoStop}%
\bibitem [{\citenamefont {{von Neuman}}\ and\ \citenamefont
  {{Wigner}}(1929)}]{Wigner}%
  \BibitemOpen
  \bibfield  {author} {\bibinfo {author} {\bibfnamefont {J.}~\bibnamefont {{von
  Neuman}}}\ and\ \bibinfo {author} {\bibfnamefont {E.}~\bibnamefont
  {{Wigner}}},\ }\bibfield  {title} {\enquote {\bibinfo {title} {{Uber
  merkw{\"u}rdige diskrete Eigenwerte. Uber das Verhalten von Eigenwerten bei
  adiabatischen Prozessen}},}\ }\href@noop {} {\bibfield  {journal} {\bibinfo
  {journal} {Physikalische Zeitschrift}\ }\textbf {\bibinfo {volume} {30}},\
  \bibinfo {pages} {467--470} (\bibinfo {year} {1929})}\BibitemShut {NoStop}%
\bibitem [{\citenamefont {Giamarchi}(2003)}]{Giamarchi}%
  \BibitemOpen
  \bibfield  {author} {\bibinfo {author} {\bibfnamefont {T.}~\bibnamefont
  {Giamarchi}},\ }\href@noop {} {\emph {\bibinfo {title} {Quantum Physics in
  One Dimension}}},\ International Series of Monographs on Physics\ (\bibinfo
  {publisher} {Clarendon Press},\ \bibinfo {year} {2003})\BibitemShut {NoStop}%
\bibitem [{\citenamefont {Gogolin}\ \emph {et~al.}(2004)\citenamefont
  {Gogolin}, \citenamefont {Nersesyan},\ and\ \citenamefont
  {Tsvelik}}]{GogolinNerseyanTsvelik}%
  \BibitemOpen
  \bibfield  {author} {\bibinfo {author} {\bibfnamefont {A.O.}\ \bibnamefont
  {Gogolin}}, \bibinfo {author} {\bibfnamefont {A.A.}\ \bibnamefont
  {Nersesyan}}, \ and\ \bibinfo {author} {\bibfnamefont {A.M.}\ \bibnamefont
  {Tsvelik}},\ }\href {https://books.google.com/books?id=BZDfFIpCoaAC} {\emph
  {\bibinfo {title} {Bosonization and Strongly Correlated Systems}}}\ (\bibinfo
   {publisher} {Cambridge University Press},\ \bibinfo {year}
  {2004})\BibitemShut {NoStop}%
\bibitem [{\citenamefont {Tsvelik}(2003)}]{Tsvelik}%
  \BibitemOpen
  \bibfield  {author} {\bibinfo {author} {\bibfnamefont {Alexei~M.}\
  \bibnamefont {Tsvelik}},\ }\href {\doibase 10.1017/CBO9780511615832} {\emph
  {\bibinfo {title} {Quantum Field Theory in Condensed Matter Physics}}},\
  \bibinfo {edition} {2nd}\ ed.\ (\bibinfo  {publisher} {Cambridge University
  Press},\ \bibinfo {year} {2003})\BibitemShut {NoStop}%
\bibitem [{\citenamefont {Abrikosov}(1983)}]{Abrisokov}%
  \BibitemOpen
  \bibfield  {author} {\bibinfo {author} {\bibfnamefont {A.}~\bibnamefont
  {Abrikosov}},\ }\bibfield  {title} {\enquote {\bibinfo {title}
  {Superconductivity in a quasi-one-dimensional metal with impurities},}\
  }\href {\doibase 10.1007/BF00682484} {\bibfield  {journal} {\bibinfo
  {journal} {Journal of Low Temperature Physics}\ }\textbf {\bibinfo {volume}
  {53}},\ \bibinfo {pages} {359--374} (\bibinfo {year} {1983})}\BibitemShut
  {NoStop}%
\bibitem [{\citenamefont {Sengupta}\ \emph {et~al.}(2001)\citenamefont
  {Sengupta}, \citenamefont {\ifmmode \check{Z}\else
  \v{Z}\fi{}uti\ifmmode~\acute{c}\else \'{c}\fi{}}, \citenamefont {Kwon},
  \citenamefont {Yakovenko},\ and\ \citenamefont {Das~Sarma}}]{Sengupta}%
  \BibitemOpen
  \bibfield  {author} {\bibinfo {author} {\bibfnamefont {K.}~\bibnamefont
  {Sengupta}}, \bibinfo {author} {\bibfnamefont {Igor}\ \bibnamefont {\ifmmode
  \check{Z}\else \v{Z}\fi{}uti\ifmmode~\acute{c}\else \'{c}\fi{}}}, \bibinfo
  {author} {\bibfnamefont {Hyok-Jon}\ \bibnamefont {Kwon}}, \bibinfo {author}
  {\bibfnamefont {Victor~M.}\ \bibnamefont {Yakovenko}}, \ and\ \bibinfo
  {author} {\bibfnamefont {S.}~\bibnamefont {Das~Sarma}},\ }\bibfield  {title}
  {\enquote {\bibinfo {title} {Midgap edge states and pairing symmetry of
  quasi-one-dimensional organic superconductors},}\ }\href {\doibase
  10.1103/PhysRevB.63.144531} {\bibfield  {journal} {\bibinfo  {journal} {Phys.
  Rev. B}\ }\textbf {\bibinfo {volume} {63}},\ \bibinfo {pages} {144531}
  (\bibinfo {year} {2001})}\BibitemShut {NoStop}%
\bibitem [{\citenamefont {Sau}\ \emph {et~al.}(2010)\citenamefont {Sau},
  \citenamefont {Tewari}, \citenamefont {Lutchyn}, \citenamefont {Stanescu},\
  and\ \citenamefont {Das~Sarma}}]{Sau}%
  \BibitemOpen
  \bibfield  {author} {\bibinfo {author} {\bibfnamefont {Jay~D.}\ \bibnamefont
  {Sau}}, \bibinfo {author} {\bibfnamefont {Sumanta}\ \bibnamefont {Tewari}},
  \bibinfo {author} {\bibfnamefont {Roman~M.}\ \bibnamefont {Lutchyn}},
  \bibinfo {author} {\bibfnamefont {Tudor~D.}\ \bibnamefont {Stanescu}}, \ and\
  \bibinfo {author} {\bibfnamefont {S.}~\bibnamefont {Das~Sarma}},\ }\bibfield
  {title} {\enquote {\bibinfo {title} {Non-abelian quantum order in
  spin-orbit-coupled semiconductors: Search for topological majorana particles
  in solid-state systems},}\ }\href {\doibase 10.1103/PhysRevB.82.214509}
  {\bibfield  {journal} {\bibinfo  {journal} {Phys. Rev. B}\ }\textbf {\bibinfo
  {volume} {82}},\ \bibinfo {pages} {214509} (\bibinfo {year}
  {2010})}\BibitemShut {NoStop}%
\bibitem [{\citenamefont {Fidkowski}\ \emph {et~al.}(2011)\citenamefont
  {Fidkowski}, \citenamefont {Lutchyn}, \citenamefont {Nayak},\ and\
  \citenamefont {Fisher}}]{Fidkowski}%
  \BibitemOpen
  \bibfield  {author} {\bibinfo {author} {\bibfnamefont {Lukasz}\ \bibnamefont
  {Fidkowski}}, \bibinfo {author} {\bibfnamefont {Roman~M.}\ \bibnamefont
  {Lutchyn}}, \bibinfo {author} {\bibfnamefont {Chetan}\ \bibnamefont {Nayak}},
  \ and\ \bibinfo {author} {\bibfnamefont {Matthew P.~A.}\ \bibnamefont
  {Fisher}},\ }\bibfield  {title} {\enquote {\bibinfo {title} {Majorana zero
  modes in one-dimensional quantum wires without long-ranged superconducting
  order},}\ }\href {\doibase 10.1103/physrevb.84.195436} {\bibfield  {journal}
  {\bibinfo  {journal} {Physical Review B}\ }\textbf {\bibinfo {volume} {84}}
  (\bibinfo {year} {2011}),\ 10.1103/physrevb.84.195436}\BibitemShut {NoStop}%
\bibitem [{\citenamefont {Fabrizio}\ and\ \citenamefont
  {Gogolin}(1995)}]{Fabrizio}%
  \BibitemOpen
  \bibfield  {author} {\bibinfo {author} {\bibfnamefont {M.}~\bibnamefont
  {Fabrizio}}\ and\ \bibinfo {author} {\bibfnamefont {Alexander~O.}\
  \bibnamefont {Gogolin}},\ }\bibfield  {title} {\enquote {\bibinfo {title}
  {Interacting one-dimensional electron gas with open boundaries},}\ }\href
  {\doibase 10.1103/PhysRevB.51.17827} {\bibfield  {journal} {\bibinfo
  {journal} {Phys. Rev. B}\ }\textbf {\bibinfo {volume} {51}},\ \bibinfo
  {pages} {17827--17841} (\bibinfo {year} {1995})}\BibitemShut {NoStop}%
\bibitem [{\citenamefont {Haldane}(1981)}]{Haldane}%
  \BibitemOpen
  \bibfield  {author} {\bibinfo {author} {\bibfnamefont {FDM}\ \bibnamefont
  {Haldane}},\ }\bibfield  {title} {\enquote {\bibinfo {title} {Effective
  harmonic-fluid approach to low-energy properties of one-dimensional quantum
  fluids},}\ }\href@noop {} {\bibfield  {journal} {\bibinfo  {journal}
  {Physical Review Letters}\ }\textbf {\bibinfo {volume} {47}},\ \bibinfo
  {pages} {1840} (\bibinfo {year} {1981})}\BibitemShut {NoStop}%
\bibitem [{\citenamefont {Delfino}\ and\ \citenamefont
  {Mussardo}(1998)}]{Delfino}%
  \BibitemOpen
  \bibfield  {author} {\bibinfo {author} {\bibfnamefont {G.}~\bibnamefont
  {Delfino}}\ and\ \bibinfo {author} {\bibfnamefont {G.}~\bibnamefont
  {Mussardo}},\ }\bibfield  {title} {\enquote {\bibinfo {title} {Non-integrable
  aspects of the multi-frequency sine-gordon model},}\ }\href {\doibase
  10.1016/s0550-3213(98)00063-7} {\bibfield  {journal} {\bibinfo  {journal}
  {Nuclear Physics B}\ }\textbf {\bibinfo {volume} {516}},\ \bibinfo {pages}
  {675–703} (\bibinfo {year} {1998})}\BibitemShut {NoStop}%
\bibitem [{\citenamefont {Fabrizio}\ \emph {et~al.}(2000)\citenamefont
  {Fabrizio}, \citenamefont {Gogolin},\ and\ \citenamefont
  {Nersesyan}}]{Fabrizio2}%
  \BibitemOpen
  \bibfield  {author} {\bibinfo {author} {\bibfnamefont {M.}~\bibnamefont
  {Fabrizio}}, \bibinfo {author} {\bibfnamefont {A.O.}\ \bibnamefont
  {Gogolin}}, \ and\ \bibinfo {author} {\bibfnamefont {A.A.}\ \bibnamefont
  {Nersesyan}},\ }\bibfield  {title} {\enquote {\bibinfo {title} {Critical
  properties of the double-frequency sine-gordon model with applications},}\
  }\href {\doibase 10.1016/s0550-3213(00)00247-9} {\bibfield  {journal}
  {\bibinfo  {journal} {Nuclear Physics B}\ }\textbf {\bibinfo {volume}
  {580}},\ \bibinfo {pages} {647–687} (\bibinfo {year} {2000})}\BibitemShut
  {NoStop}%
\bibitem [{\citenamefont {{Fateev}}(1996)}]{Fateev}%
  \BibitemOpen
  \bibfield  {author} {\bibinfo {author} {\bibfnamefont {V.~A.}\ \bibnamefont
  {{Fateev}}},\ }\bibfield  {title} {\enquote {\bibinfo {title} {{The sigma
  model (dual) representation for a two-parameter family of integrable quantum
  field theories}},}\ }\href {\doibase 10.1016/0550-3213(96)00256-8} {\bibfield
   {journal} {\bibinfo  {journal} {Nuclear Physics B}\ }\textbf {\bibinfo
  {volume} {473}},\ \bibinfo {pages} {509--538} (\bibinfo {year}
  {1996})}\BibitemShut {NoStop}%
\bibitem [{\citenamefont {{Bukhvostov}}\ and\ \citenamefont
  {{Lipatov}}(1981)}]{BL}%
  \BibitemOpen
  \bibfield  {author} {\bibinfo {author} {\bibfnamefont {A.~P.}\ \bibnamefont
  {{Bukhvostov}}}\ and\ \bibinfo {author} {\bibfnamefont {L.~N.}\ \bibnamefont
  {{Lipatov}}},\ }\bibfield  {title} {\enquote {\bibinfo {title}
  {{Instanton-anti-instanton interaction in the O(3) non-linear {$\sigma$}
  model and an exactly soluble fermion theory}},}\ }\href {\doibase
  10.1016/0550-3213(81)90157-7} {\bibfield  {journal} {\bibinfo  {journal}
  {Nuclear Physics B}\ }\textbf {\bibinfo {volume} {180}},\ \bibinfo {pages}
  {116--140} (\bibinfo {year} {1981})}\BibitemShut {NoStop}%
\bibitem [{\citenamefont {Lesage}\ \emph {et~al.}(1998)\citenamefont {Lesage},
  \citenamefont {Saleur},\ and\ \citenamefont {Simonetti}}]{Lesage}%
  \BibitemOpen
  \bibfield  {author} {\bibinfo {author} {\bibfnamefont {F.}~\bibnamefont
  {Lesage}}, \bibinfo {author} {\bibfnamefont {H.}~\bibnamefont {Saleur}}, \
  and\ \bibinfo {author} {\bibfnamefont {P.}~\bibnamefont {Simonetti}},\
  }\bibfield  {title} {\enquote {\bibinfo {title} {Tunneling in quantum wires
  ii: A line of ir fixed points},}\ }\href {\doibase 10.1103/physrevb.57.4694}
  {\bibfield  {journal} {\bibinfo  {journal} {Physical Review B}\ }\textbf
  {\bibinfo {volume} {57}},\ \bibinfo {pages} {4694–4707} (\bibinfo {year}
  {1998})}\BibitemShut {NoStop}%
\bibitem [{\citenamefont {Saleur}(1999)}]{SaleurBL}%
  \BibitemOpen
  \bibfield  {author} {\bibinfo {author} {\bibfnamefont {H}~\bibnamefont
  {Saleur}},\ }\bibfield  {title} {\enquote {\bibinfo {title} {The long delayed
  solution of the bukhvostov-lipatov model},}\ }\href {\doibase
  10.1088/0305-4470/32/18/102} {\bibfield  {journal} {\bibinfo  {journal}
  {Journal of Physics A: Mathematical and General}\ }\textbf {\bibinfo {volume}
  {32}},\ \bibinfo {pages} {L207–L213} (\bibinfo {year} {1999})}\BibitemShut
  {NoStop}%
\bibitem [{\citenamefont {Timm}(2012)}]{Tim}%
  \BibitemOpen
  \bibfield  {author} {\bibinfo {author} {\bibfnamefont {Carsten}\ \bibnamefont
  {Timm}},\ }\bibfield  {title} {\enquote {\bibinfo {title} {Transport through
  a quantum spin hall quantum dot},}\ }\href {\doibase
  10.1103/PhysRevB.86.155456} {\bibfield  {journal} {\bibinfo  {journal} {Phys.
  Rev. B}\ }\textbf {\bibinfo {volume} {86}},\ \bibinfo {pages} {155456}
  (\bibinfo {year} {2012})}\BibitemShut {NoStop}%
\bibitem [{\citenamefont {{Takahashi}}(1999)}]{Takahashi}%
  \BibitemOpen
  \bibfield  {author} {\bibinfo {author} {\bibfnamefont {M.}~\bibnamefont
  {{Takahashi}}},\ }\href@noop {} {\emph {\bibinfo {title} {Thermodynamics of
  One-Dimensional Solvable Models, by Minoru Takahashi, Cambridge, UK:
  Cambridge University Press, 1999}}}\ (\bibinfo {year} {1999})\BibitemShut
  {NoStop}%
\bibitem [{\citenamefont {{Korepin}}\ \emph {et~al.}(1993)\citenamefont
  {{Korepin}}, \citenamefont {{Bogoliubov}},\ and\ \citenamefont
  {{Izergin}}}]{KorepinBook}%
  \BibitemOpen
  \bibfield  {author} {\bibinfo {author} {\bibfnamefont {V.~E.}\ \bibnamefont
  {{Korepin}}}, \bibinfo {author} {\bibfnamefont {N.~M.}\ \bibnamefont
  {{Bogoliubov}}}, \ and\ \bibinfo {author} {\bibfnamefont {A.~G.}\
  \bibnamefont {{Izergin}}},\ }\href@noop {} {\emph {\bibinfo {title} {Quantum
  Inverse Scattering Method and Correlation Functions}}}\ (\bibinfo
  {publisher} {Cambridge University Press},\ \bibinfo {year} {1993})\ p.\
  \bibinfo {pages} {575}\BibitemShut {NoStop}%
\bibitem [{\citenamefont {{Gaudin}}\ and\ \citenamefont
  {{Caux}}(2014)}]{Gaudin}%
  \BibitemOpen
  \bibfield  {author} {\bibinfo {author} {\bibfnamefont {M.}~\bibnamefont
  {{Gaudin}}}\ and\ \bibinfo {author} {\bibfnamefont {J.-S.}\ \bibnamefont
  {{Caux}}},\ }\href@noop {} {\emph {\bibinfo {title} {The Bethe Wavefunction,
  by Michel Gaudin , Translated by Jean-S{\'e}bastien Caux, Cambridge, UK:
  Cambridge University Press, 2014}}}\ (\bibinfo {year} {2014})\BibitemShut
  {NoStop}%
\bibitem [{\citenamefont {Andrei}(1992, cond-mat/9408101)}]{4lectures}%
  \BibitemOpen
  \bibfield  {author} {\bibinfo {author} {\bibfnamefont {Natan}\ \bibnamefont
  {Andrei}},\ }\bibfield  {title} {\enquote {\bibinfo {title} {Integrable
  models in condensed matter physics},}\ }in\ \href@noop {} {\emph {\bibinfo
  {booktitle} {Series on Modern Condensed Matter Physics - Vol. 6, Lecture
  Notes of ICTP Summer Course}}},\ \bibinfo {editor} {edited by\ \bibinfo
  {editor} {\bibfnamefont {G.~Morandi}\ \bibnamefont {S.~Lundquist}}\ and\
  \bibinfo {editor} {\bibfnamefont {Yu}~\bibnamefont {Lu}}}\ (\bibinfo
  {publisher} {World Scientific},\ \bibinfo {year} {1992, cond-mat/9408101})\
  pp.\ \bibinfo {pages} {458 -- 551}\BibitemShut {NoStop}%
\bibitem [{Note1()}]{Note1}%
  \BibitemOpen
  \bibinfo {note} {A discussion of the different choices of regularization is
  presented in [\protect \citenum {Roy}]. See [\protect \citenum
  {BergknoffThacker}] and [\protect \citenum {Korepinmtm}] for two different
  choices in the sine-Gordon model with different expressions for the phase
  shift in terms of the interaction strength. The regularization used in the
  text corresponds to that which is used in [\protect \citenum
  {BergknoffThacker}]. Also see [\protect \citenum {Camacho}] for a discussion
  in other models and a comparison with bosonization}\BibitemShut {NoStop}%
\bibitem [{\citenamefont {Rylands}\ and\ \citenamefont
  {Andrei}(2018)}]{RylandsAndrei}%
  \BibitemOpen
  \bibfield  {author} {\bibinfo {author} {\bibfnamefont {Colin}\ \bibnamefont
  {Rylands}}\ and\ \bibinfo {author} {\bibfnamefont {Natan}\ \bibnamefont
  {Andrei}},\ }\bibfield  {title} {\enquote {\bibinfo {title} {Quantum dot in
  interacting environments},}\ }\href {\doibase 10.1103/PhysRevB.97.155426}
  {\bibfield  {journal} {\bibinfo  {journal} {Phys. Rev. B}\ }\textbf {\bibinfo
  {volume} {97}},\ \bibinfo {pages} {155426} (\bibinfo {year}
  {2018})}\BibitemShut {NoStop}%
\bibitem [{\citenamefont {Gangadharaiah}\ \emph {et~al.}(2011)\citenamefont
  {Gangadharaiah}, \citenamefont {Braunecker}, \citenamefont {Simon},\ and\
  \citenamefont {Loss}}]{Gangadharaiah}%
  \BibitemOpen
  \bibfield  {author} {\bibinfo {author} {\bibfnamefont {Suhas}\ \bibnamefont
  {Gangadharaiah}}, \bibinfo {author} {\bibfnamefont {Bernd}\ \bibnamefont
  {Braunecker}}, \bibinfo {author} {\bibfnamefont {Pascal}\ \bibnamefont
  {Simon}}, \ and\ \bibinfo {author} {\bibfnamefont {Daniel}\ \bibnamefont
  {Loss}},\ }\bibfield  {title} {\enquote {\bibinfo {title} {Majorana edge
  states in interacting one-dimensional systems},}\ }\href {\doibase
  10.1103/physrevlett.107.036801} {\bibfield  {journal} {\bibinfo  {journal}
  {Physical Review Letters}\ }\textbf {\bibinfo {volume} {107}} (\bibinfo
  {year} {2011}),\ 10.1103/physrevlett.107.036801}\BibitemShut {NoStop}%
\bibitem [{\citenamefont {Stoudenmire}\ \emph {et~al.}(2011)\citenamefont
  {Stoudenmire}, \citenamefont {Alicea}, \citenamefont {Starykh},\ and\
  \citenamefont {Fisher}}]{Stoudenmire}%
  \BibitemOpen
  \bibfield  {author} {\bibinfo {author} {\bibfnamefont {E.~M.}\ \bibnamefont
  {Stoudenmire}}, \bibinfo {author} {\bibfnamefont {Jason}\ \bibnamefont
  {Alicea}}, \bibinfo {author} {\bibfnamefont {Oleg~A.}\ \bibnamefont
  {Starykh}}, \ and\ \bibinfo {author} {\bibfnamefont {Matthew~P.A.}\
  \bibnamefont {Fisher}},\ }\bibfield  {title} {\enquote {\bibinfo {title}
  {Interaction effects in topological superconducting wires supporting majorana
  fermions},}\ }\href {\doibase 10.1103/PhysRevB.84.014503} {\bibfield
  {journal} {\bibinfo  {journal} {Phys. Rev. B}\ }\textbf {\bibinfo {volume}
  {84}},\ \bibinfo {pages} {014503} (\bibinfo {year} {2011})}\BibitemShut
  {NoStop}%
\bibitem [{Note2()}]{Note2}%
  \BibitemOpen
  \bibinfo {note} {This change of the right hand boundary condition results in
  a shift of the boundary contribution to the energy density, however the
  boundary energy comes from a redistribution of the Fermi sea which decays
  exponentially away from the Fermi level. At the Fermi level the modification
  amounts to an additional constant phase shift acquired by particle reflecting
  off the boundary which will not provide a shift in the boundary energy. As
  such in the thermodynamic limit the modification results in a negligible
  change in the energy. The bound state energy is unaffected by this
  shift.}\BibitemShut {Stop}%
\bibitem [{\citenamefont {{Sklyanin}}(1988)}]{Sklyannin}%
  \BibitemOpen
  \bibfield  {author} {\bibinfo {author} {\bibfnamefont {E.~K.}\ \bibnamefont
  {{Sklyanin}}},\ }\bibfield  {title} {\enquote {\bibinfo {title} {{Boundary
  conditions for integrable quantum systems}},}\ }\href {\doibase
  10.1088/0305-4470/21/10/015} {\bibfield  {journal} {\bibinfo  {journal}
  {Journal of Physics A Mathematical General}\ }\textbf {\bibinfo {volume}
  {21}},\ \bibinfo {pages} {2375--2389} (\bibinfo {year} {1988})}\BibitemShut
  {NoStop}%
\bibitem [{\citenamefont {Fendley}\ and\ \citenamefont
  {Saleur}(1998)}]{FendleySaleur}%
  \BibitemOpen
  \bibfield  {author} {\bibinfo {author} {\bibfnamefont {P.}~\bibnamefont
  {Fendley}}\ and\ \bibinfo {author} {\bibfnamefont {H.}~\bibnamefont
  {Saleur}},\ }\bibfield  {title} {\enquote {\bibinfo {title} {Self-duality in
  quantum impurity problems},}\ }\href {\doibase 10.1103/physrevlett.81.2518}
  {\bibfield  {journal} {\bibinfo  {journal} {Physical Review Letters}\
  }\textbf {\bibinfo {volume} {81}},\ \bibinfo {pages} {2518–2521} (\bibinfo
  {year} {1998})}\BibitemShut {NoStop}%
\bibitem [{Note3()}]{Note3}%
  \BibitemOpen
  \bibinfo {note} {In order to compare the energy of the states with and
  without the bound states present the cutoff must be the same for both. The
  cutoff for the interacting case, $\Lambda _g$, is determined using $D=\DOTSI
  \intop \ilimits@ _0^{\Lambda _g}\protect \mathrm {d}\theta \protect \tmspace
  +\thinmuskip {.1667em}\rho (\theta )$ in the absence of the bound states. It
  is then held fixed and the energy with and without the bound states is
  calculated and compared. Alternatively we could have determined the cutoff
  with the bound states included already with the same result.}\BibitemShut
  {Stop}%
\bibitem [{\citenamefont {Maurand}\ \emph {et~al.}(2012)\citenamefont
  {Maurand}, \citenamefont {Meng}, \citenamefont {Bonet}, \citenamefont
  {Florens}, \citenamefont {Marty},\ and\ \citenamefont
  {Wernsdorfer}}]{KondoPRX}%
  \BibitemOpen
  \bibfield  {author} {\bibinfo {author} {\bibfnamefont {Romain}\ \bibnamefont
  {Maurand}}, \bibinfo {author} {\bibfnamefont {Tobias}\ \bibnamefont {Meng}},
  \bibinfo {author} {\bibfnamefont {Edgar}\ \bibnamefont {Bonet}}, \bibinfo
  {author} {\bibfnamefont {Serge}\ \bibnamefont {Florens}}, \bibinfo {author}
  {\bibfnamefont {La\"etitia}\ \bibnamefont {Marty}}, \ and\ \bibinfo {author}
  {\bibfnamefont {Wolfgang}\ \bibnamefont {Wernsdorfer}},\ }\bibfield  {title}
  {\enquote {\bibinfo {title} {First-order
  $0\mathrm{\text{\ensuremath{-}}}\ensuremath{\pi}$ quantum phase transition in
  the kondo regime of a superconducting carbon-nanotube quantum dot},}\ }\href
  {\doibase 10.1103/PhysRevX.2.011009} {\bibfield  {journal} {\bibinfo
  {journal} {Phys. Rev. X}\ }\textbf {\bibinfo {volume} {2}},\ \bibinfo {pages}
  {011009} (\bibinfo {year} {2012})}\BibitemShut {NoStop}%
\bibitem [{\citenamefont {{Sakurai}}(1970)}]{Sakurai}%
  \BibitemOpen
  \bibfield  {author} {\bibinfo {author} {\bibfnamefont {A.}~\bibnamefont
  {{Sakurai}}},\ }\bibfield  {title} {\enquote {\bibinfo {title} {{Comments on
  Superconductors with Magnetic Impurities}},}\ }\href {\doibase
  10.1143/PTP.44.1472} {\bibfield  {journal} {\bibinfo  {journal} {Progress of
  Theoretical Physics}\ }\textbf {\bibinfo {volume} {44}},\ \bibinfo {pages}
  {1472--1476} (\bibinfo {year} {1970})}\BibitemShut {NoStop}%
\bibitem [{\citenamefont {Sakai}\ \emph {et~al.}(1993)\citenamefont {Sakai},
  \citenamefont {Shimizu}, \citenamefont {Shiba},\ and\ \citenamefont
  {Satori}}]{SakaiShimizu}%
  \BibitemOpen
  \bibfield  {author} {\bibinfo {author} {\bibfnamefont {Osamu}\ \bibnamefont
  {Sakai}}, \bibinfo {author} {\bibfnamefont {Yukihiro}\ \bibnamefont
  {Shimizu}}, \bibinfo {author} {\bibfnamefont {Hiroyuki}\ \bibnamefont
  {Shiba}}, \ and\ \bibinfo {author} {\bibfnamefont {Koji}\ \bibnamefont
  {Satori}},\ }\bibfield  {title} {\enquote {\bibinfo {title} {Numerical
  renormalization group study of magnetic impurities in superconductors. ii.
  dynamical excitation spectra and spatial variation of the order parameter},}\
  }\href {\doibase 10.1143/JPSJ.62.3181} {\bibfield  {journal} {\bibinfo
  {journal} {Journal of the Physical Society of Japan}\ }\textbf {\bibinfo
  {volume} {62}},\ \bibinfo {pages} {3181--3197} (\bibinfo {year} {1993})},\
  \Eprint {http://arxiv.org/abs/https://doi.org/10.1143/JPSJ.62.3181}
  {https://doi.org/10.1143/JPSJ.62.3181} \BibitemShut {NoStop}%
\bibitem [{\citenamefont {M\"uller-Hartmann}\ and\ \citenamefont
  {Zittartz}(1971)}]{Zitt}%
  \BibitemOpen
  \bibfield  {author} {\bibinfo {author} {\bibfnamefont {E.}~\bibnamefont
  {M\"uller-Hartmann}}\ and\ \bibinfo {author} {\bibfnamefont {J.}~\bibnamefont
  {Zittartz}},\ }\bibfield  {title} {\enquote {\bibinfo {title} {Kondo effect
  in superconductors},}\ }\href {\doibase 10.1103/PhysRevLett.26.428}
  {\bibfield  {journal} {\bibinfo  {journal} {Phys. Rev. Lett.}\ }\textbf
  {\bibinfo {volume} {26}},\ \bibinfo {pages} {428--432} (\bibinfo {year}
  {1971})}\BibitemShut {NoStop}%
\bibitem [{\citenamefont {Bortolin}\ \emph {et~al.}(2019)\citenamefont
  {Bortolin}, \citenamefont {Iucci},\ and\ \citenamefont {Lobos}}]{Iucci}%
  \BibitemOpen
  \bibfield  {author} {\bibinfo {author} {\bibfnamefont {Tom\'as}\ \bibnamefont
  {Bortolin}}, \bibinfo {author} {\bibfnamefont {An\'{\i}bal}\ \bibnamefont
  {Iucci}}, \ and\ \bibinfo {author} {\bibfnamefont {Alejandro~M.}\
  \bibnamefont {Lobos}},\ }\bibfield  {title} {\enquote {\bibinfo {title}
  {Quantum phase diagram of shiba impurities from bosonization},}\ }\href
  {\doibase 10.1103/PhysRevB.100.155111} {\bibfield  {journal} {\bibinfo
  {journal} {Phys. Rev. B}\ }\textbf {\bibinfo {volume} {100}},\ \bibinfo
  {pages} {155111} (\bibinfo {year} {2019})}\BibitemShut {NoStop}%
\bibitem [{\citenamefont {Roy}(1993)}]{Roy}%
  \BibitemOpen
  \bibfield  {author} {\bibinfo {author} {\bibfnamefont {C.~L.}\ \bibnamefont
  {Roy}},\ }\bibfield  {title} {\enquote {\bibinfo {title} {Boundary conditions
  across a \ensuremath{\delta}-function potential in the one-dimensional dirac
  equation},}\ }\href {\doibase 10.1103/PhysRevA.47.3417} {\bibfield  {journal}
  {\bibinfo  {journal} {Phys. Rev. A}\ }\textbf {\bibinfo {volume} {47}},\
  \bibinfo {pages} {3417--3419} (\bibinfo {year} {1993})}\BibitemShut {NoStop}%
\bibitem [{\citenamefont {Bergknoff}\ and\ \citenamefont
  {Thacker}(1979)}]{BergknoffThacker}%
  \BibitemOpen
  \bibfield  {author} {\bibinfo {author} {\bibfnamefont {H.}~\bibnamefont
  {Bergknoff}}\ and\ \bibinfo {author} {\bibfnamefont {H.~B.}\ \bibnamefont
  {Thacker}},\ }\bibfield  {title} {\enquote {\bibinfo {title} {Structure and
  solution of the massive thirring model},}\ }\href {\doibase
  10.1103/PhysRevD.19.3666} {\bibfield  {journal} {\bibinfo  {journal} {Phys.
  Rev. D}\ }\textbf {\bibinfo {volume} {19}},\ \bibinfo {pages} {3666--3681}
  (\bibinfo {year} {1979})}\BibitemShut {NoStop}%
\bibitem [{\citenamefont {Korepin}(1980)}]{Korepinmtm}%
  \BibitemOpen
  \bibfield  {author} {\bibinfo {author} {\bibfnamefont {V.~E.}\ \bibnamefont
  {Korepin}},\ }\bibfield  {title} {\enquote {\bibinfo {title} {The mass
  spectrum and thes matrix of the massive thirring model in the repulsive
  case},}\ }\href {\doibase 10.1007/BF01212824} {\bibfield  {journal} {\bibinfo
   {journal} {Communications in Mathematical Physics}\ }\textbf {\bibinfo
  {volume} {76}},\ \bibinfo {pages} {165--176} (\bibinfo {year}
  {1980})}\BibitemShut {NoStop}%
\bibitem [{\citenamefont {Camacho}\ \emph {et~al.}(2019)\citenamefont
  {Camacho}, \citenamefont {Schmitteckert},\ and\ \citenamefont
  {Carr}}]{Camacho}%
  \BibitemOpen
  \bibfield  {author} {\bibinfo {author} {\bibfnamefont {Gonzalo}\ \bibnamefont
  {Camacho}}, \bibinfo {author} {\bibfnamefont {Peter}\ \bibnamefont
  {Schmitteckert}}, \ and\ \bibinfo {author} {\bibfnamefont {Sam~T.}\
  \bibnamefont {Carr}},\ }\bibfield  {title} {\enquote {\bibinfo {title} {Exact
  equilibrium results in the interacting resonant level model},}\ }\href
  {\doibase 10.1103/PhysRevB.99.085122} {\bibfield  {journal} {\bibinfo
  {journal} {Phys. Rev. B}\ }\textbf {\bibinfo {volume} {99}},\ \bibinfo
  {pages} {085122} (\bibinfo {year} {2019})}\BibitemShut {NoStop}%
\end{thebibliography}%
 \end{document}